\documentclass[prd,twocolumn,10pt,
superscriptaddress,
nofootinbib,
]{revtex4-2}

\usepackage{aas_macros}
\usepackage{graphicx}
\usepackage{dcolumn}
\usepackage{bm}
\usepackage[utf8]{inputenc}
\usepackage{amsmath,amssymb}
\usepackage{graphicx}
\usepackage[dvipsnames]{xcolor}

\usepackage{hyperref}
\usepackage{cprotect}
\usepackage{verbatim}
\usepackage{multirow}
\usepackage{orcidlink}
\usepackage{comment}
\usepackage{siunitx}
\usepackage{xspace}
\usepackage[normalem]{ulem}

\hypersetup{
 colorlinks  = true, 
 urlcolor   = RedOrange, 
 linkcolor  = RedOrange, 
 citecolor  = RedOrange 
}
\usepackage[capitalise]{cleveref}

\newcommand{\SNR}{\mathrm{SNR}}

\newcommand{\mm}{\mathcal{M}}
\newcommand{\pfp}{p_{\mathrm{FP}}}

\newcommand{\lnB}{\ln \mathcal{B}}
\newcommand{\MSun}{M_\odot}
\newcommand{\Msun}{\MSun}
\newcommand{\ymax}{y_{\mathrm{max}}}
\newcommand{\thetalens}{\theta_{\mathrm{lens}}}
\newcommand{\thetasrc}{\theta_{\mathrm{src}}}
\newcommand{\HU}{\mathcal{H}_\mathrm{U}}
\newcommand{\HL}{\mathcal{H}_\mathrm{L}}
\newcommand{\hU}{h_\mathrm{U}}
\newcommand{\hL}{h_\mathrm{L}}
\newcommand{\Bthr}{\mathcal{B}_\mathrm{thresh}}

\newcommand{\ip}[2]{\left(#1\middle|#2\right)}  
\newcommand{\dhp}{\delta h_\perp}
\newcommand{\ML}{\mathrm{ML}}
\newcommand{\true}{\mathrm{true}}

\newcommand{\Dth}{D_\theta}
\newcommand{\DL}{D_\lambda}

\DeclareMathOperator{\CDF}{CDF}

\begin{document}

\preprint{APS/123-QED}

\title{Diffraction of gravitational waves by extended dark objects}

\author{Nikita Blinov\orcidlink{0000-0002-2845-961X}}
\email{nblinov@yorku.ca}
\affiliation{Department of Physics and Astronomy, York University, Toronto, ON M3J 1P3, Canada}

\author{J. Leo Kim\orcidlink{0000-0001-8699-834X}}
\email{jlkim@yorku.ca}
\affiliation{Department of Physics and Astronomy, York University, Toronto, ON M3J 1P3, Canada}

\date{\today}

\begin{abstract}
Many dark sector models predict the formation of dark objects, such as solitons and dark stars, which can be searched for through their gravitational lensing of gravitational waves (GWs).
Although these objects vary widely in size and compactness, most GW lensing studies have focused on point-like lenses. The finite size of an extended lens is important when it exceeds its Einstein radius.
We evaluate the frequency-dependent effects on GW waveforms from extended lenses and demonstrate that they can be probed with current and future GW data. While finite-size effects can weaken the bounds on the fraction of dark matter in these objects compared to the point-like limit, current GW data provides complementary constraints to other gravitational tests for lens masses between $10^2\Msun$ and $10^5\Msun$; future LIGO-Virgo-KAGRA data and next-generation observatories will surpass their sensitivity.
As expected, we find that the finite-size effects are important when the lens radius exceeds $\sim \text{few} \times 10^{6}R_\odot (M_L/100 M_\odot)^{1/2}$.
Interpreting these constraints within specific models (a simple dissipative dark sector and axion stars), we show that current and future observations probe significant portions of the microphysical parameter space. 
\end{abstract}

\maketitle

\section{Introduction\label{sec:level1}}

The growing catalogue of gravitational wave (GW) events provides a powerful tomographic probe of the dark matter (DM) distribution in the Universe. 
DM structures can imprint many different signals on GW observations through gravitational lensing. For example, DM halos with masses in excess of $10^8\Msun$ can produce significant time delays between different paths around the lens, resulting in multiple GW ``images'' of a single binary coalescence event in ground-based interferometers~\cite{Takahashi:2003ix}. The images can also be magnified, which can affect the inference of the merger parameters~\cite{Wang:1996as,Dai:2016igl,Dai:2018enj,Oguri:2018muv} (see also~\cite{Grespan:2023cpa} for a recent review). These phenomena are well known from the gravitational lensing of light, where geometric optics applies. No conclusive evidence for strong lensing has emerged from LIGO-Virgo-KAGRA (LVK) searches so far~\cite{LIGOScientific:2021izm,LIGOScientific:2023bwz,LIGOScientific:2025cwb}. 

Unlike astrophysical sources of electromagnetic radiation, binary mergers produce coherent GWs with wavelengths $\lambda$ that can be comparable to the Schwarzschild radius $R_S$ of astrophysical objects. In this regime, wave optics effects become important: the GWs are diffracted, developing frequency-dependent amplitude modulations and phase shifts~\cite{Nakamura:1997sw,Takahashi:2003ix}. For the LIGO band, $\lambda \sim \qtyrange{3e2}{3e4}{km}$, so these effects are relevant for lens masses as low as $\sim 10 \Msun$, providing an exciting opportunity for studying relatively light DM substructures~\cite{Jung:2017flg, Liao:2020hnx, Basak:2021ten,Wang:2021lij, Urrutia:2021qak}. To date, the LVK Collaboration has not found evidence for GW lensing in this regime either~\cite{LIGOScientific:2021izm,LIGOScientific:2023bwz} (although see \cite{Goyal:2025eqo}).

Small-scale DM objects are natural predictions of many dark sector models; their mass and compactness are highly model-dependent and therefore can span several orders of magnitude. 
For example, dissipative dark sectors can contain DM particles with long-range interactions similar to electromagnetism, which enable radiative cooling and the formation of dark analogues of stars \cite{Kouvaris:2015rea, Maselli:2017vfi, Buckley:2017ttd, Ghalsasi:2017jna, Chang:2018bgx, Gurian:2022nbx, Bramante:2023ddr, Bramante:2024pyc, Buckley:2024eoe, Litterer:2025quq, aDMstars}. Other examples include bosonic solitons which occur in scalar field models of DM, such as axions \cite{Marsh:2015wka, Marsh:2015xka, Arvanitaki:2019rax, Chadha-Day:2021szb, Chang:2024fol, Hardy:2026mkp}. Non-gravitational interactions between DM and the Standard Model (SM) enable a variety of constraints on the abundance of such objects~\cite{Curtin:2019ngc, Curtin:2019lhm, Bhoonah:2020dzs, Bai:2023mfi, Kaplan:2024dsn, Bramante:2024hbr, Alonso-Alvarez:2024ypq, Bramante:2026wzh, Picker:2026nuu}. However, in the absence of observational evidence for these interactions, it is important to consider purely gravitational probes, such as microlensing \cite{Paczynski, EROS-2:2006ryy, Niikura:2017zjd}, dynamical heating \cite{Graham:2023unf}, accretion \cite{Ali-Haimoud:2016mbv, Agius:2024ecw}, and GW emission \cite{Banks:2023eym, BetancourtKamenetskaia:2026cyf}.
Although most studies have explored these in the context of point-like objects, such as primordial black holes (PBHs), recent works have generalized these constraints by modelling finite-size effects~\cite{Croon:2020wpr, Croon:2020ouk, Bai:2020jfm, DeRocco:2023hij, Kim:2025gck, Graham:2025opw, Osuna:2026dzj, Croon:2026uqm}. The diffractive lensing of GWs provides a complementary gravitational window onto these objects, which we explore in this work.

Regardless of their microphysical origin, \emph{extended} dark objects can differ significantly in their lensing properties compared to the point lens~\cite{Nakamura:1997sw,Takahashi:2003ix,Jung:2017flg, Liao:2020hnx, Basak:2021ten,Wang:2021lij, Urrutia:2021qak}. Naively, the shallower gravitational potentials of extended lenses lead to weaker effects. As we will see, this intuition is mostly correct, except in the vicinity of caustics, where the extended lens can produce lensing signatures that are absent for a point lens of the same mass.

Wave optics effects due to extended lenses have been considered recently~\cite{Dai:2018enj, Fairbairn:2022xln,Villarrubia-Rojo:2024xcj,Ephremidze:2026era}. Our primary goal here is to derive constraints from merger events in the currently available LVK data, using the Gravitational Wave Transient Catalogs (GWTC-1.0 to 5.0~\cite{GWTC1,GWTC2-1,GWTC3,GWTC4,GWTC5}), and to interpret them in concrete particle physics models. We also improve on aspects of previous analyses by using realistic \verb|Phenom| waveforms~\cite{Ajith:2007kx} and noise spectra, and by employing a simple, easily interpretable Bayesian criterion to quantify the detectability of lensing~\cite{Owen:1995tm, Lindblom:2008cm,Savastano:2023spl}. Throughout this work, we study the single-lens regime relevant for LVK frequency band observations; at lower frequencies (such as those accessible with LISA), stochastic lensing due to multiple perturbers along the line of sight may be relevant~\cite{Oguri:2020ldf,Zumalacarregui:2024ocb,Kim:2025njb}.

\begin{figure*}
\includegraphics[width=0.85\textwidth]{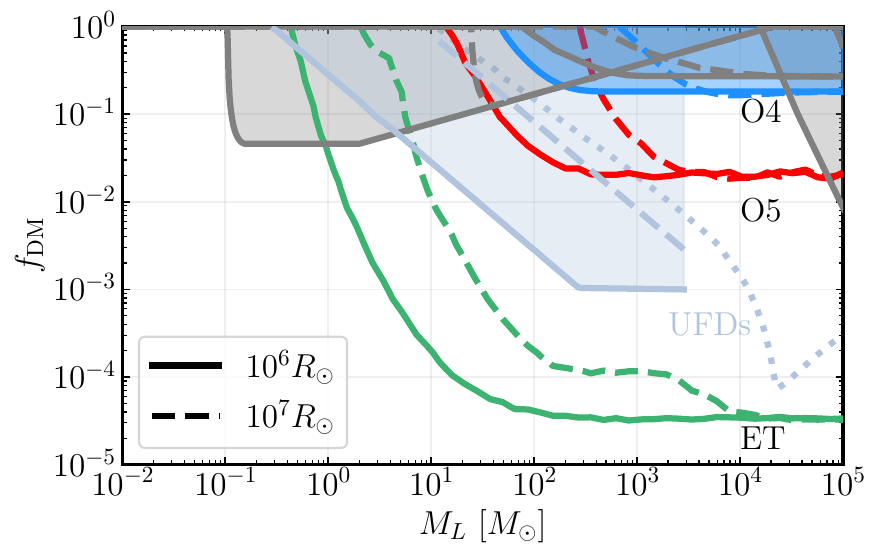}
\caption{\label{fig:f_DM_const_R_comparisons} Current and projected upper limits on the extended dark object fraction $f_\mathrm{DM}$ as a function of the compact object mass $M_L$, assuming a monochromatic distribution for the compact objects. The curves labelled O4 are from the total number of GW events detected up to the fourth observing run of the LVK network, whereas LIGO A+ (O5) and Einstein Telescope (ET) are projected curves assuming an observation time of 1 year. Note that for $R \lesssim 10^6 R_\odot$, the constraints for GW lensing are nearly identical to the point lens constraints. For comparison, we also show existing constraints for an extended object in the gray shaded regions (see Ref.~\cite{Croon:2024jhd} for a repository of such constraints). From left to right, these include microlensing of the caustic-crossing star Icarus \cite{Croon:2025yfj}, binary evaporation \cite{Ramirez:2022mys}, and the cosmic microwave background (CMB) \cite{Croon:2024rmw}. We have also overlaid in light-blue/gray curves constraints from ultrafaint dwarfs (UFDs). The solid and dashed light-blue/gray curves are from wide binaries in the UFD Bo\"otes I, which was studied assuming a $\rho \propto r^{-3/2}$ density profile for the extended dark objects \cite{Olea-Romacho:2026pgn}. The region enclosed by the dotted curve shows relevant constraints on $f_{\rm DM}$ from the heating of ultrafaint dwarfs \cite{Graham:2023unf, Graham:2025opw}, which were studied assuming point-like perturbers.}
\end{figure*}

The rest of the paper is organized as follows. In \cref{sec:lensing} we review the GW lensing formalism. Next, we compute the expected number of lensing events at current and next-generation GW detectors in \cref{sec:detectability} and use it to constrain the fraction of DM in lenses of various sizes and project the sensitivity of future observations; a summary of our results is shown in \cref{fig:f_DM_const_R_comparisons}. Then in \cref{sec:compact_objects}, we demonstrate how these bounds map onto the microphysical parameter space of specific models which motivate the existence of extended DM lenses: a simple, asymmetric, dissipative dark sector \cite{Chang:2018bgx, Bramante:2023ddr, Bramante:2024pyc, Litterer:2025quq} and axion stars (see, e.g., \cite{Tkachev:1991ka,Seidel:1993zk} and recent reviews \cite{Braaten:2019knj,Visinelli:2021uve}). Finally, we conclude in \cref{sec:conclusion}. The appendices provide derivations for our statistical methodology (\cref{sec:stats}), validate various approximations (\cref{sec:mm_minimization,sec:ligo_validation}), and translate other constraints to our choice of the density profile (\cref{sec:icarus_uniform}). Throughout this work we use Planck units with $G = c = \hbar = 1$, except in \cref{sec:adm} where we display $G$ explicitly.

\section{Gravitational wave lensing} \label{sec:lensing}
Lensed GWs are described by the dimensionless amplification factor $F(f)$ which relates the Fourier transforms of the lensed $\tilde{h}_{\rm L}$ and unlensed $\tilde{h}_{\rm U}$ waveforms:
\begin{align}
  F(f) = \frac{\tilde{h}_{\rm L}(f)}{\tilde{h}_{\rm U}(f)}\,.
\end{align}
In general, $F$ is complex, and therefore (despite its name) describes both the frequency-dependent amplification and phase shift of the GW. $F$ is obtained by solving the wave equation in the eikonal and thin lens approximations. The result is given in terms of the Kirchhoff-Fresnel diffraction integral~\cite{1999PThPS.133..137N, Takahashi:2003ix, Leung:2023lmq}:\footnote{This result can also be derived from scattering theory -- see \cite{CarrilloGonzalez:2025gqm}.}
\begin{align}
  F(f) = \frac{D_S \xi_0^2}{D_L D_{LS}} \frac{f}{i} \int d^2 \bm{x} ~ e^{2\pi i f t_d (\bm{x}, \bm{y})}, \label{eq:F_full}
\end{align}
where $\bm{x} = \bm{\xi}/\xi_0$ and $\bm{y} = D_L\bm{\eta}/(\xi_0 D_S) $ are dimensionless forms for the impact parameter $\bm{\xi}$ and the source position vector $\bm{\eta}$; $D_L$, $D_S$, and $D_{LS}$ are the observer-lens, observer-source, and lens-source angular diameter distances, respectively; and $\xi_0$ is an arbitrary normalization constant, which for the lenses considered in this work is given by
\begin{align}
  \xi_0 = R_E= D_L \theta_E,
\end{align}
where $R_E$ and $\theta_E$ are the Einstein radius and Einstein angle, respectively. The delay time $t_d$ is given by
\begin{align}
  t_d (\mathbf{x}, \mathbf{y}) = \frac{w}{2\pi} \phi(\mathbf{x}, \mathbf{y}) \,,
\end{align}
where the dimensionless parameter $w$ for a given lens redshift $z_L$ is
\begin{align}
  w \equiv 2 \pi f (1+z_L) \frac{D_S \xi_0^2}{D_L D_{LS}} \,. \label{eq:w_def}
\end{align}
Note that the factor $D_S\xi_0^2/(D_L D_{LS})$ is independent of the distances $D_{L,S,LS}$ and it is generally proportional to the Schwarzschild radius; for example it equals $4M_L$ for the point lens (see \cref{eq:w_point_lens} below). Note that for a sufficiently massive lens the time delay can exceed the duration of LVK events, so that different paths around the lens appear as distinct mergers; this regime requires a different search strategy. We therefore focus on $M_{L} \leq 10^5 M_\odot$ to avoid this regime~\cite{Takahashi:2003ix}. 

For a single axisymmetric lens, the coordinates $\bm{x}$ can be centred on the lens and rotated such that the Fermat potential is given by 
\begin{align}
  \phi(x,y) = \frac{1}{2} (x - y)^2 - \psi(x) - \phi_m(y), 
\end{align}
with $y>0$ specifying the position of the source~\cite{Schneider:1992bmb}. $\phi_m(y)$ is chosen so that the minimum arrival time is zero, i.e. $\min_{x} t_d(x,y) = 0$. 
The deflection potential $\psi(x)$ and therefore $\phi_m$ depend on the detailed mass distribution of the lens. 
This means that the diffraction integral \cref{eq:F_full} can be written in a dimensionless form as
\begin{align}
  F(f) = \frac{w}{2 \pi i} \int d^2 \bm{x} \exp[i w \phi(x,y)]. \label{eq:F_dimless}
\end{align}
The regime $w\sim1$ marks the transition between wave and geometric optics (GO). The GO limit corresponds to $w\gg 1$, where the wavelength of the GW is much smaller than the Schwarzschild radius of the lens. Thus the diffraction integral can be approximated by a sum over the stationary points of $t_d$, such that 
\begin{align}
  F(f) \approx \sum_j |\mu (x_j)|^{1/2} e^{iw \phi(x_j,y) - i a _j\pi},
  \label{eq:F_geo_approx}
\end{align}
where $x_j$ are critical points of the Fermat potential, $\nabla \phi = 0$, and $a_j = 0, 1/2$ or $1$ if the point is a minimum, saddle, or maximum, respectively. 
For an axisymmetric lens, the magnification factor $\mu_j = \mu(x_j)$ of an image $j$ is given by~\cite{Schneider:1992bmb}
\begin{align}
  \mu_j^{-1} = \left(1- \frac{\psi'(x_j)}{x_j} \right) \left( 1 - \psi''(x_j) \right) \,,
\end{align}
where primes indicate derivatives with respect to the radial coordinate $x$.
\cref{eq:F_geo_approx} is useful for evaluating waveform overlap integrals which include a wide range of frequencies (and therefore $w$) as we describe below.

The angular diameter distances are computed via
\begin{align}
  D_{X} = \frac{1}{1+z_{X}} \int_0^{z_{X}} \frac{dz'}{H(z')},
\end{align}
where $X=S$ or $L$ for the source and lens respectively, and $H(z)$ is the Hubble parameter. The lens-source distance is then given by $D_{LS} \equiv D_S - (1+z_L)/(1+z_S)D_L$. Throughout this work, we use Planck18 values for cosmological parameters, namely an expansion rate of $H_0 = 67.4$ km/s/Mpc, $\Omega_m = 0.315$, $\Omega_\Lambda = 1 - \Omega_m$~\cite{Planck:2018vyg}. 

In the following sections we will model extended dark objects as uniform density spheres of various masses and radii. The deflection potential outside of such a lens is the same as that of a point mass, so we will first review the point mass regime. This similarity will also enable certain numerical simplifications when we scan over the lens parameter space. 
\subsection{Point lens}
\label{sec:point_lens}
The point lens is a useful description of any object as long as its physical size $R$ is much smaller than the Einstein radius:
\begin{equation}
R \ll R_E \sim \qty{9e3}{AU}\sqrt{\frac{M_L}{10\Msun} \frac{\unit{Gpc}}{D_{LS} D_L/D_S}},
\label{eq:point_lens_limit}
\end{equation}
where we used the Einstein angle \cite{Takahashi:2003ix}
\begin{equation}
  \theta_E = \sqrt{4M_L\frac{D_{LS}}{D_L D_S}}.
  \label{eq:point_lens_einstein_angle}  
\end{equation} 
We will see how the condition in \cref{eq:point_lens_limit} emerges for an extended lens. The dimensionless parameter $w$ in this case can be written as
\begin{align}
  w = 8 \pi f M_{Lz},\; M_{Lz} \equiv (1+z_L) M_L,
  \label{eq:w_point_lens}
\end{align}
while the lensing potential is given by
\begin{align}
  \psi(x) &= \ln x.
\end{align}
The corresponding Fermat potential has two extrema at
\begin{align}
  x_\pm &= \frac{1}{2} (\pm y + \sqrt{y^2 + 4})\,,
\end{align}
and therefore two images in the geometric $w\gg 1$ limit; the minimum 
time delay occurs at $x_+$, so that 
\begin{align}
  \phi_m(y) &= \frac{1}{2} (x_+ - y)^2 - \ln x_+ .
\end{align}
Substituting these into \cref{eq:F_dimless} yields an analytic result for the diffraction integral~\cite{Peters:1974gj,Takahashi:2003ix}
\begin{align}
  F_\mathrm{point}(f) = &\exp \left[ \frac{\pi w}{4} + \frac{iw}{2} \left(\ln \frac{w}{2} - 2 \phi_m(y) \right) \right] \nonumber \\
  &\times \Gamma\left(1 - \frac{iw}{2} \right) {}_1F_1 \left( \frac{iw}{2}, 1, \frac{iw}{2} y^2 \right) ,
  \label{eq:point_lens_diffraction_integral}
\end{align}
where ${}_1F_1$ is the confluent hypergeometric function and $\Gamma$ is the usual gamma function. 
In what follows, we will compare the amplification from a point lens with that from an extended lenses. 

\subsection{Uniform density sphere}
\label{sec:uniform_ball_lens}
A uniform density sphere is often used to approximate the density of collapsed structures in simplified dissipative dark matter models \cite{Kouvaris:2015rea, Maselli:2017vfi, Buckley:2017ttd, Chang:2018bgx, Gurian:2022nbx, Bramante:2023ddr, Bramante:2024pyc, Buckley:2024eoe, Litterer:2025quq} and of solitonic cores of axion substructure \cite{Marsh:2015wka, Chang:2024fol}. For an object of mass $M_L$ and physical radius $R$, the density profile is
\begin{align}
  \rho(r) = \frac{3M_L}{4 \pi R^3} \Theta(R-r),
\end{align}
where $\Theta$ is the Heaviside function. The lensing potential is given by \cite{Fairbairn:2022xln}
\begin{align}
  \label{eq:ball_pot}
  \psi(x) = 
  \begin{cases}
    \frac{1}{3} \sqrt{1 - \frac{x^2}{b^2}} ( \frac{x^2}{b^2} - 4) + \ln (b + \sqrt{b^2 - x^2}), &x \leq b \\
    \ln x, &x > b,
  \end{cases}
\end{align} 
where $b = R/R_E$, in which $R_E$ is the Einstein radius of the object (given by \cref{eq:point_lens_einstein_angle} as for a point mass). Since the lensing potential is the same as for a point lens for $x > b$, we only need to compute the Fresnel integral over $x < b$, and subtract the contribution from the point lens. In other words, to find the amplification factor we compute
\begin{align}
  F_\mathrm{ball}(f) = F_\mathrm{point}(f) + \frac{w}{2\pi i} \int_0^b d^2 \mathbf{x} ~ &\left(e^{i w \phi_\mathrm{ball}} - e^{i w \phi_\mathrm{point}} \right),
  \label{eq:ball_amplification_decomposition}
\end{align}
where $\phi_\mathrm{ball}$ and $\phi_\mathrm{point}$ are the Fermat potentials for the uniform sphere and the point lens, respectively. 
This is a useful decomposition because $F_\mathrm{point}(f)$ is known analytically (\cref{eq:point_lens_diffraction_integral}) and an efficient implementation is provided by the Gravitational Lensing of Waves (\verb|GLoW|) package \cite{Villarrubia-Rojo:2024xcj}. The second term in \cref{eq:ball_amplification_decomposition} is computed numerically; we found that it is more stable over a wide range of lens parameters, compared to performing the entire integral in \cref{eq:F_dimless} numerically owing to the finite range of integration. The interferometer sensitivity calculations in \cref{sec:detectability} involve integrals over the appropriate frequency band, so $w \propto f$ can be large, leading to rapid oscillations even in the finite integral part of \cref{eq:ball_amplification_decomposition}. In order to avoid numerical quadrature convergence issues at $w\gg 1$, we use the geometric optics limit, \cref{eq:F_geo_approx}, for $w \geq w_\mathrm{thresh} \gg 1$. 
We have validated our calculation against other works (e.g. Refs.~\cite{Fairbairn:2022xln, Caliskan:2023zqm}).\footnote{We note that although \texttt{GLoW} has a built-in class for the constant density sphere, we found that it does not always find all of the stationary points needed to apply the GO approximation. We therefore use it only for validation of our computation for the amplification factor.} We take $w_{\rm thresh} \sim 150$, such that we evaluate the diffraction integral numerically well into the GO regime, before the integration becomes unstable due to rapid oscillations.

\begin{figure}
\includegraphics[width=\columnwidth]{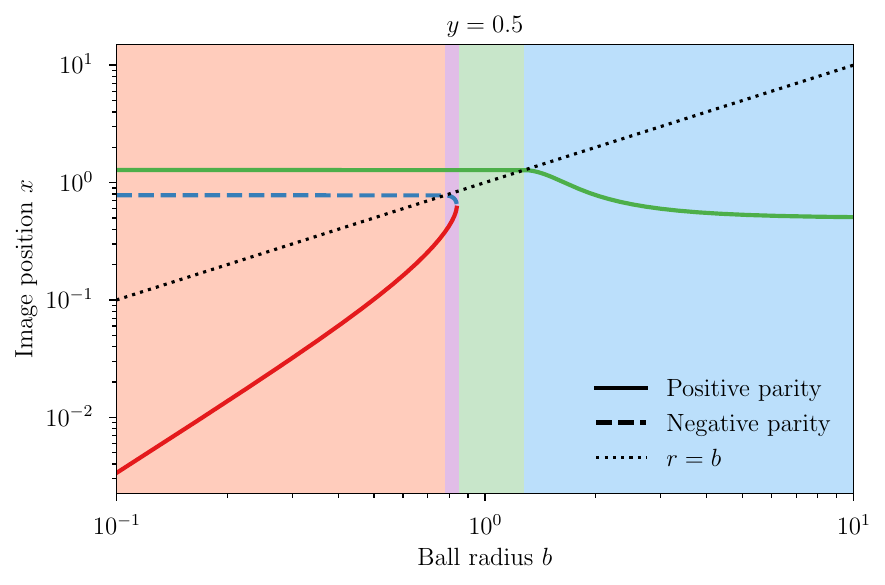}
\caption{\label{fig:image_tracks} Image positions for the uniform sphere lens as a function of the sphere radius $b$. Solid and dashed lines indicate the positive and negative parity images, respectively, while the dotted line is $x = b$. For $x < b$ ($x>b$) the images are inside (outside) the lens. The four regimes discussed in the text are shaded in different colours.}
\end{figure}

\begin{figure}
\includegraphics[width=\columnwidth]{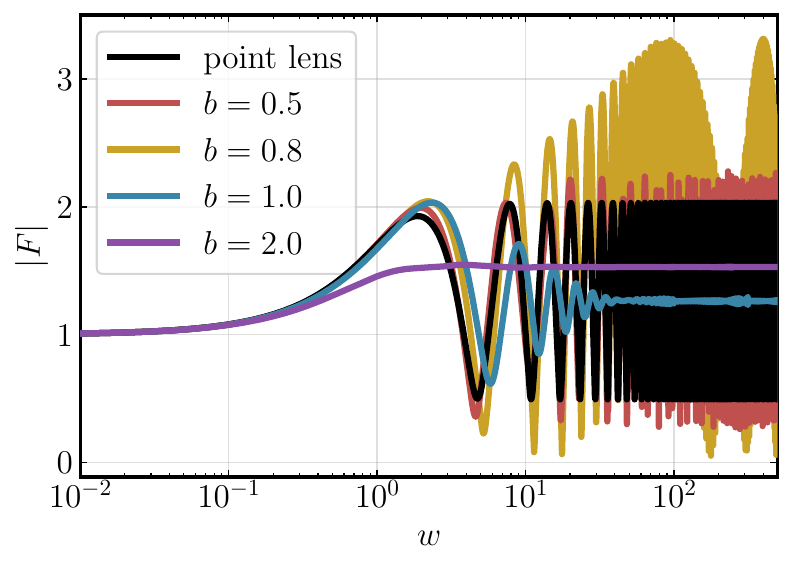}
\caption{\label{fig:F_comparison} Amplification factor $F(w)$ assuming a fixed impact parameter of $y = 0.5$ for a point-like lens (black) and for $b=0.5, 0.8, 1.0$, and $2.0$ (red, yellow, blue, purple, respectively). Each of the various $b$-values corresponds to a different regime for $b$ and $y$, leading to a different number of extremal paths (see main text). The amplification factor for the point lens was computed using \texttt{GLoW} \cite{Villarrubia-Rojo:2024xcj}.}
\end{figure}

Next we consider the amplification factors for the uniform sphere lens for various values of $b$ and compare them to the point lens. The behaviour of the diffraction integral can be qualitatively understood in terms of the number of stationary points of the integrand in \cref{eq:F_dimless} (even if we are not in the strict geometric optics limit); the stationary points are the solutions of 
\begin{align}
  \frac{\partial \psi}{\partial x} - x \pm y = 0, \label{eq:Fermat_min}
\end{align}
where $\psi$ is the lensing potential given in \cref{eq:ball_pot}.
Following \cite{Fairbairn:2022xln}, we find four different cases depending on the compactness of the object (through $b$) and the impact parameter $y$. In \cref{fig:image_tracks} we illustrate these regimes by plotting the image positions as a function of $b$ at a fixed value of $y$. The corresponding behaviours of the diffraction integral are shown in \cref{fig:F_comparison}.
The different regimes are:
\begin{enumerate}
  \item \textbf{point-like regime}. If $0 < b \leq 1$ and $y \leq (1-b^2)/b$, there are three roots of \cref{eq:Fermat_min}, with two at $x> b$ and one at $x \leq b$. For $b \to 0$ the amplification factor approaches that for a point lens. This is clear in the geometric optics regime, where the $x<b$ stationary point contributes only at $\mathcal{O}(b^2)$ to the diffraction integral; the dominant contributions therefore come from the two point-lens-like images at $x > b$. Thus $b\ll 1$ corresponds to the point-lens limit, justifying the condition in \cref{eq:point_lens_limit}. 
  For $b\lesssim 1$ there can be a slight increase in amplification when compared to the point lens case, cf. the red curve in \cref{fig:F_comparison}. This regime was called ``Type I'' in \cite{Fairbairn:2022xln}.
  \item \textbf{near-caustic regime}. As we increase $b$ at fixed $y$, eventually one of the external roots moves inside the lens, which occurs when $x_- = b$. For a narrow interval in $b$ two images with opposite parity exist at $x\lesssim b$. As $b$ increases further they annihilate at a critical line as the source crosses the corresponding caustic. This caustic is a unique feature of the uniform density sphere. In the geometric optics limit, this narrow region of parameter space is associated with large magnifications; however, even in the wave optics regime the diffraction integral is enhanced. This regime was indicated as a grey band in Fig. 2 of \cite{Fairbairn:2022xln}.
  
  \item \textbf{weak lensing regime}. For $x_+ > b\gtrsim x_-$, there is a single image outside of the lens. The magnitude of the amplification factor is close to 1; it oscillates briefly before it asymptotes to a small overall amplification in the geometric optics limit. Both features can be seen in the blue curve in \cref{fig:F_comparison}. This regime was called ``Type II'' in \cite{Fairbairn:2022xln}.
  \item \textbf{diffuse regime}. When $b$ crosses $x_+$, the remaining image is inside the lens. The magnitude of the diffraction integral asymptotes to a fixed amplification with very minimal oscillation. This can be seen in \cref{fig:F_comparison} as the purple curve. This regime was called ``Type III'' in \cite{Fairbairn:2022xln}.
\end{enumerate}

\section{Detectability} \label{sec:detectability}
\subsection{Detection criterion}
\label{sec:det_crit}
\begin{figure*}
\includegraphics[width=\textwidth]{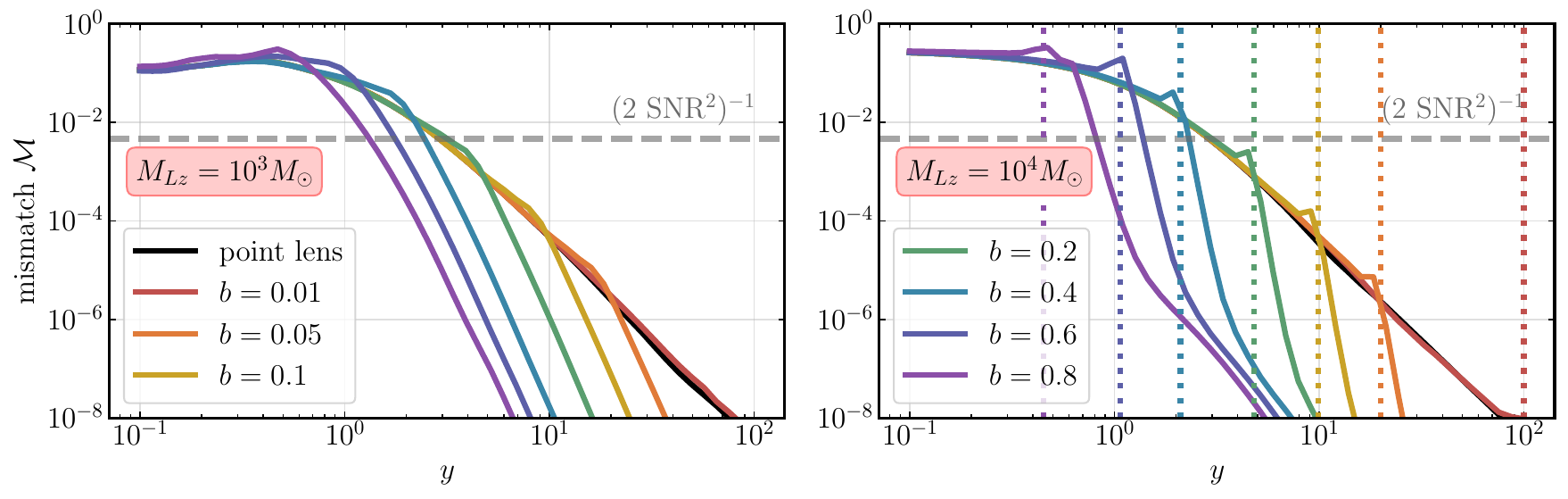}
\caption{\label{fig:y_max_example} Mismatch $\mm$ as a function of $y$ assuming a point-like lens (black line), or a uniform sphere lens (coloured lines) with radii ranging from $b=0.01$ to $b=0.8$. The left panel corresponds to a redshifted lens mass of $M_{Lz} = 10^3 M_\odot$ whereas the right panel is for $M_{Lz} = 10^4 M_\odot$. As an example, we have chosen source parameters from \texttt{GW240109\_050431} in the GWTC-4.0 catalogue -- an event with $m_1 = 28.8 M_\odot$, $m_2 = 18.1 M_\odot$, and $z_S = 0.29$, with $\SNR = 10.4$. The grey dashed line is the threshold $1/(2\,\mathrm{SNR}^2)$. The vertical dotted lines indicate the location of the geometric-optics caustic, where the number of images changes.}
\end{figure*} 
In this section we quantify the detection probability of diffractive deformations of GWs and interpret it in terms of the sensitivity to the DM fraction in lenses. Previous works have used various $\chi^2$-type statistics to compare lensed and unlensed waveforms~\cite{Jung:2017flg,Fairbairn:2022xln,Villarrubia-Rojo:2024xcj}, while LVK and related works generally employ Bayesian techniques~\cite{Basak:2021ten,LIGOScientific:2021izm,LIGOScientific:2023bwz,LIGOScientific:2025pvj}. Here we relate the two approaches, showing how the simple $\chi^2$ framework emerges in the high signal-to-noise ratio (SNR) limit and that it is able to reproduce the sensitivity calculations of a full Bayesian analysis for point-like lenses. This serves as validation for our approach, which we then apply to re-interpret existing measurements and make sensitivity projections for extended lenses in the following sections. 

From the Bayesian perspective, for each detected GW event we want to compare two hypotheses: $\HU$ -- the event is unlensed, or $\HL$ -- the event is lensed. The Bayes factor~\cite{2019PASA...36...10T}
\begin{equation}
  \mathcal{B}(\thetalens) = \frac{\int d \theta p(d|\HL(\theta, \thetalens))\pi(\theta,\thetalens)}{\int d \theta p(d|\HU(\theta))\pi(\theta)} \,,
  \label{eq:bayes_factor}
\end{equation}
expresses the preference of $\HL$ over $\HU$ given data $d$. In the above expression, $p(d | \mathcal{H}_i)$ is the likelihood (probability of data given hypothesis $\mathcal{H}_i$) and $
\pi$ is the prior probability distribution of parameters of $\mathcal{H}_i$. Their product is proportional to the posterior probability distribution according to Bayes' theorem. Since $\mathcal{B}$ is a random variable, the diffraction detection probability is the total probability for $\mathcal{B}$ to exceed some pre-determined threshold $\Bthr$.
We will consider reasonable values of $\Bthr$ below. 

For each event, the data stream is 
\begin{equation}
  d(t) = h(t) + n(t)\,,
\end{equation}
where $h(t)$ is the true GW waveform and the noise $n(t)$ is assumed to be a stationary Gaussian random process. The latter assumption means that the likelihoods are Gaussian~\cite{Maggiore:2007ulw}: 
\begin{equation}
  p(d | \mathcal{H}_i(\theta)) \propto \exp\left(-\frac{1}{2} ( d - h_i(\theta) | d - h_i(\theta) ) \right)
\end{equation}
where $h_i$ is the GW strain corresponding to $\mathcal{H}_i$ and $\ip{\cdot}{\cdot}$ denotes the noise-weighted inner product
\begin{align}
  (h_1|h_2) \equiv 4 \text{Re} \int_{f_\mathrm{min}}^{f_\mathrm{max}} df~ \frac{\tilde{h}_1(f)^* \tilde{h}_2(f)}{S_n(f)}.
  \label{eq:overlap}
\end{align}
The quantity $S_n$ in \cref{eq:overlap} is the noise power spectral density (PSD) for a given detector. We obtain noise PSDs for various detectors and GW waveforms using the \verb|PyCBC| package \cite{alex_nitz_2024_10473621}, utilizing the \verb|IMRPhenomD| waveform model. The frequency integration range in \cref{eq:overlap} depends on the specific GW detector. The lower bound is $f_\mathrm{min} = 10$ Hz for current LVK data and $f_\mathrm{min} = 1$ Hz for the LVK O5 and Einstein Telescope (ET) projections. For the upper bounds we take either the maximum modelled frequency for the \verb|PyCBC| waveforms or the maximum detectable frequency from the PSD for a given detector. 

In the high SNR limit, one can use the Laplace approximation~\cite{Kass:1995loi} to perform the integrals in \cref{eq:bayes_factor} to show that $2\ln \mathcal{B}$ follows a $\chi^2$ distribution under different noise realizations. We follow Ref.~\cite{Vallisneri:2012qq} and provide the details in \cref{sec:stats}. In particular we find that when the data contains a lensed signal 
\begin{equation}
2\ln \mathcal{B} - c_L \sim \chi^2_3((\delta h_\perp | \delta h_\perp )) \,,
\label{eq:lnB_under_HL}
\end{equation}
where $\chi^2_3$ is the non-central $\chi^2$ distribution with 3 degrees of freedom, and $c_L$ is an $\mathcal{O}(1)$ Occam correction. 
The three degrees of freedom are the extra parameters of the extended lens model ($M_L$, $R$, $y$; for a point lens this would be reduced by 1).

Ref.~\cite{Vallisneri:2012qq} computed the non-centrality parameter to be
\begin{equation}
  (\delta h_\perp | \delta h_\perp ) \approx 2\; \SNR^2 \mm \, ,
  \label{eq:non_centrality}
\end{equation}
where $\mm$ is the minimized mismatch between the lensed and unlensed waveforms~\cite{Owen:1995tm, Lindblom:2008cm, Tambalo:2022wlm, Savastano:2023spl, Singh:2025uvp, Lin:2025mpx, LIGOScientific:2025cwb}\footnote{In this work we take $\mm$ as the mismatch following standard convention, not to be confused with the chirp mass, for which we always use the redshifted quantity $\mathcal{M}_z$.}
\begin{equation}
    \mm = \min_{\theta} \left[1 - \frac{(\hU(\theta)|\hL)}{ \sqrt{(\hU(\theta)|\hU(\theta))} \sqrt{(\hL|\hL)}}\right]
  \label{eq:mismatch} \,.
\end{equation}
The minimization is over the unlensed waveform parameters. Here $\hL$ is the lensed waveform at the maximum-posterior point, i.e. $\hL \approx h$ when the data is lensed. This means that $\mm$ is a function of the true source and lens parameters. In writing \cref{eq:non_centrality} we took 
\begin{equation}
\SNR^2 \equiv (\hL|\hL) \approx (\hU|\hU),
\label{eq:SNR}
\end{equation}
i.e. lensed and unlensed waveforms give similar SNR, an approximation that holds to $\mathcal{O}(1/\SNR^4)$~\cite{Vallisneri:2012qq} near the detectability threshold $(\delta h_\perp | \delta h_\perp ) \sim 1$.
 Larger non-centrality parameters correspond to diffractive deformations of GW waveforms that are more detectable; physically the minimization in \cref{eq:mismatch} means that detectability requires the lensing effect to exceed noise \emph{and} that it cannot be absorbed by adjusting some parameters of the unlensed waveform. Note that $\mm$ is insensitive to waveform amplitudes -- it measures only the shape mismatch. 

The result in \cref{eq:lnB_under_HL} holds when all of the integrands in \cref{eq:bayes_factor} are approximately Gaussian. This means that $ (\delta h_\perp | \delta h_\perp )$ cannot be small. Therefore, as a rough guideline we adopt $(\delta h_\perp | \delta h_\perp ) > 1$ as a validity condition for what follows, which is similar to the Lindblom criterion~\cite{Owen:1995tm, Lindblom:2008cm}. We use this condition to define the maximum impact parameter $\ymax$ beyond which the approximations leading to \cref{eq:lnB_under_HL} are not valid:
\begin{equation}
  2\; \SNR^2 \mm(M_L, R, \ymax, \thetasrc) = 1 \,,
  \label{eq:ymax_def}
\end{equation}
where $\thetasrc$ is a shorthand for all of the GW source parameters.

\cref{fig:y_max_example} shows the mismatch $\mm$ as a function of $y$ for several radii $b$ and redshifted lens masses $M_{Lz} = 10^3 M_\odot$ (left panel) and $10^4 M_\odot$ (right panel). For source parameters we took the event \verb|GW240109_050431| from GWTC-4.0 with $m_1 = 28.8 M_\odot$, $m_2 = 18.1 M_\odot$, $z_S = 0.29$, and $\SNR = 10.4$. As expected, for values of $b$ which are close to 0 (in the sense of \cref{eq:point_lens_limit}), the mismatch converges to that of a point lens. However, for large values of $y$, the mismatch from the uniform sphere lens will eventually deviate from that of a point lens due to the lens entering the single-image regime. For each value of $b$, the dotted vertical lines indicate the boundary at which $y = |1-b^2|/b$, where the number of extremal paths changes from 1 to 3, resulting in large fluctuations in the diffraction integral $F$. This, in turn, leads to an increase in the mismatch. 
The grey dashed line in \cref{fig:y_max_example} shows $\mm_\mathrm{thresh} = (2\;\SNR^{2})^{-1}$; the intersection of this line with the mismatch is $\ymax$. A simple approximation for the diffraction lensing probability follows: $\epsilon_{\rm lens} = \Theta(\ymax-y)$. This can be used to estimate the optical depth below. This approximation misses the fact that even for $y < \ymax$, noise fluctuations can occasionally overwhelm the lensing effect -- i.e., in reality $\epsilon_{\rm lens} < 1$. We account for this next. 

To estimate the detection probability we need to set a threshold $\Bthr$. A low threshold increases the diffraction detection probability \emph{and} the false positive probability $\pfp$ -- the chance that a noise fluctuation is mis-identified as a lensed event. This is the usual tradeoff between the false alarm and signal rate. In what follows we take $\pfp = 1/N$, where $N$ is the number of GW events in the analysis, so that on average one false positive is expected across the whole catalogue. In the framework described above $\pfp$ is the total probability for $\mathcal{B}$ to exceed the threshold under the null hypothesis -- the data does not contain a lensed event. Using the same Laplace approximation under the null gives
\begin{equation}
  2\ln \mathcal{B} - c_U \sim \chi_1^2 \,,
  \label{eq:lnB_under_null}
\end{equation}
where $\chi_1^2$ is the central $\chi^2$ distribution with one degree of freedom, and $c_U$ is an $\mathcal{O}(1)$ Occam correction.\footnote{\label{fn:stat}The one degree of freedom corresponds to one of the remaining lens model degrees of freedom that remains identifiable under the null hypothesis. For example, if we choose it to be $M_L$, then we expect the posterior probability to be peaked at $M_L = 0$; at this point the posteriors of the other parameters $R$ and $y$ are decidedly non-Gaussian since they have no effect on the waveform when $M_L$ is small. The fact that $M_L>0$ means that the actual distribution of $2\ln \mathcal{B}$ is not exactly $\chi^2_1$ but rather a ``Chernoff mixture''~\cite{Algeri:2019lah}; we use $\chi^2_1$ for simplicity, and because it is a broader distribution which yields more conservative values of $\Bthr$.} This enables us to relate the threshold to the desired $\pfp$:
\begin{equation}
\pfp = 1- \CDF_{\chi_1^2}(2\ln \Bthr - c_U) \,.
\end{equation}
Combining this result with \cref{eq:lnB_under_HL} enables us to evaluate the lensing detection probability/efficiency
\begin{equation}
\epsilon_{\rm lens} = 1 - \mathrm{CDF}_{\chi^2_{3}\left( (\delta h_\perp | \delta h_\perp ) \right)}\Big(\mathrm{CDF}^{-1}_{\chi^2_1}\big(1 - p_{\rm FP}\big) + \Delta c\Big) \,,
\label{eq:pdet_from_pfp}
\end{equation}
where $\Delta c = c_U - c_L$. Note that $\epsilon_{\rm lens}$ inherits the dependence on the true lens parameters ($M_L$, $R$, $y$) that live in $(\delta h_\perp | \delta h_\perp )$. The Occam penalty $\Delta c$ can in principle be computed from the priors and Fisher matrices evaluated as described in \cref{sec:stats}; at that point however, one might as well do a signal injection study to calibrate $\epsilon_{\rm lens}$ without relying on the Laplace approximation. We therefore retain $\Delta c$ as a parameter which encodes the larger parameter space of the lensed model, noting that $\Delta c > 0$ effectively increases the threshold for detection of a lens. In particular we naively expect $\Delta c$ to be larger for an extended lens model than for a point-like one, since the former has a larger parameter space. We have found that taking $\Delta c = 1.2$ gives reasonable agreement between the point lens and uniform ball constraints at large $M_L$ and fixed $R$, where the expected numbers of lensed events for the two models should coincide.

Accurate evaluation of the detection probability requires minimizing the mismatch, \cref{eq:mismatch}, over all of the parameters of $\hU$: binary masses $m_{1,2}$, source distance and angular position, orbital inclination, spins, orbital eccentricity, coalescence time $t_c$ and phase $\phi_0$. While this is less computationally intensive than a direct calculation of \cref{eq:bayes_factor}, the multidimensional optimization is still prohibitive for the parameter scans we will carry out. For this reason, in what follows we minimize only over $t_c$ and $\phi_0$. Variations of some of the remaining parameters can partially replicate certain features of diffraction, while others cannot. For example, misalignment between spin and orbital angular momentum of the merging binary leads to precession and results in modulation of the waveform~\cite{Mishra:2023ddt}. Therefore, minimizing over $t_c$ and $\phi_0$ only produces an upper bound on the mismatch and therefore somewhat optimistic lensing sensitivity. 
We quantify the quality of this approximation in \cref{sec:mm_minimization} across a broad range of parameter space, finding that it is off by at most $50\%$; there we also test the impact of such mismatch reduction on the derived constraints on the DM fraction. 
In practice, the minimization over $\phi_0$ and $t_c$ is performed using the \verb|PyCBC| function \verb|match|. 

\subsection{Lensing probability}
Finally, we evaluate the probability that a GW is lensed as it propagates from the source to us in terms of the optical depth $\tau$~\cite{Peebles:1994xt}
\begin{align}
  P_L(\theta) = 1 - (1 - p_{\rm FP}) e^{-\tau(\theta)}, \label{eq:gen_prob}
\end{align}
where 
\begin{align}
  \tau(\theta) = \int\limits_0^{z_S} dz_L \int\limits_0^{\ymax} dy\; \epsilon_{\rm lens}(\theta) ~\frac{n(z_L)[2 \pi R_E(\theta)^2 y]}{(1+z_L) H(z_L)}. \label{eq:optical_depth}
\end{align}
Note that we cut off the $y$ integration at $\ymax$ to ensure that our statistical approximation remains valid. 
Assuming that the DM objects are uniformly distributed, their number density is given by
\begin{align}
  n(z_L) = (1+z_L)^3 \frac{f_{\rm DM} \rho_\mathrm{DM}}{M_L},
\end{align}
where $\rho_{\rm DM}$ is the present-day background DM density and $f_{\rm DM}$ is the fraction of dark matter inside of these objects, assuming a monochromatic distribution for the lenses. Note that the explicit dependence on the lens mass $M_L$ drops out in the product $n(z_L) R_E^2$. Therefore all $M_L$ dependence is through $y_\mathrm{max}$ and $\epsilon_{\rm lens}$. A simple upper bound on $\tau$ can be obtained by setting $\epsilon_{\rm lens} = 1$, in which case the $y$ integral of the quantity in brackets in \cref{eq:optical_depth} is the effective lensing cross-section $\sigma = \pi R_E^2 y_\mathrm{max}^2$.

The optical depth for any given lensing event is small, so $P_L \approx \tau$. For the median source redshift in GWTC-5 of $z_S \approx 0.38$ and taking $\ymax \sim 1$ (see \cref{fig:y_max_example}) and $\epsilon_{\rm lens} = 1$, \cref{eq:optical_depth} evaluates to $\tau \sim 0.01f_{\mathrm{DM}}$. This suggests that at least hundreds of observed GW events are needed to provide useful constraints on the fraction of DM in these dense objects. 

\subsection{Events at LIGO-Virgo-KAGRA}

\begin{figure}
\includegraphics[width=\columnwidth]{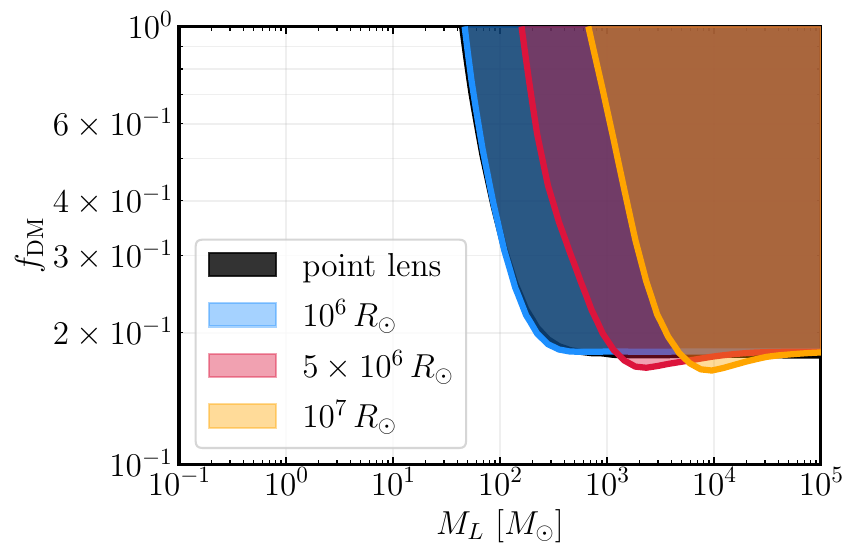}\caption{\label{fig:f_DM_constraints_LIGO_combined}LVK O4 constraints on $f_{\rm DM}$ as a function of the lens mass $M_L$, assuming the extended dark objects have fixed radii. The black contour (mostly hidden behind the blue region) corresponds to the point-like lens limit, appropriate for black holes. The coloured contours are for several values of the radius $R$. We have assumed an Occam difference factor of $\Delta c = 1.2$ for the extended lenses.}
\end{figure}

With the current $\sim 280$ events detected in the LVK network with available source parameter information, we can use the non-observation of lensing events to place constraints on the compact objects. Using \cref{eq:gen_prob}, the number of events expected at LVK is given by
\begin{align}
  N_\mathrm{events} = \sum_{i} P_L (\theta_i),
\end{align}
where $i$ runs over all of the GW events in GWTC-5.0.
To place limits on $f_{\rm DM}$ at the 95\% confidence level due to the null observation of lensing events in the GW catalogues, we require that $N_{\rm events} \leq 3$, assuming that no events have been confidently identified within the data. Note that this procedure of setting a limit on $f_{\rm DM}$ differs from the Bayesian method used by LVK~\cite{Basak:2021ten, LIGOScientific:2021izm, LIGOScientific:2023bwz} but is consistent with other theoretical work~(see, e.g., \cite{Jung:2017flg,Fairbairn:2022xln,Villarrubia-Rojo:2024xcj}). In \cref{sec:ligo_validation} we show that our simplified approach closely matches the O3 LVK result~\cite{LIGOScientific:2023bwz} for one of their prior choices. We therefore conclude that our procedure provides a reasonable approximation to a full analysis.

In \cref{fig:f_DM_constraints_LIGO_combined} we show constraints from the non-observation of lensing events in the LVK data up to GWTC-5.0 on $f_\mathrm{DM}$ as a function of $M_L$, assuming that the extended dark objects have a fixed size. Along with the point lens constraints, we show the sensitivity curves for $R = 10^6 R_\odot$, $5 \times 10^6 R_\odot$, and $10^7 R_\odot$. For objects with radii $R \lesssim 10^6 R_\odot$, the densities are sufficiently high that their lensing effects are similar to those of a point lens (see \cref{eq:point_lens_limit}), and so the constraints for the $10^6 R_\odot$ case are very similar to those of a point lens. As one would expect, however, the constraints on the extended lenses are relaxed when assuming larger sizes at a fixed mass due to the rapid decrease in the density. With the current LVK data available up to GWTC-5.0, we are not sensitive to lenses with sizes $R \gtrsim 10^{8} R_\odot$ for masses $M_L \lesssim 10^5 \Msun$.

\subsection{Events at future detectors}

We now consider possible constraints on the population of compact objects from future detectors. First, the next set of LVK upgrades before O5, called A+, is expected to improve the sensitivity to DM fraction in point lenses from $\mathcal{O}(10^{-1})$ to $\mathcal{O}(10^{-2})$ \cite{Basak:2021ten,LIGOScientific:2023bwz}. In addition, next-generation ground-based observatories such as the European Einstein Telescope \cite{Branchesi:2023mws} and the US-based Cosmic Explorer (CE) \cite{Reitze:2019iox} will significantly improve the sensitivity to GW events, and consequently the number of expected GW lensing events. We derive projections for the near-term A+ LVK configuration, and the longer-term ET observatory. 

To forecast the number of lensing events we first estimate the total number of expected GW events as~\cite{Urrutia:2021qak, Fairbairn:2022xln}
\begin{align}
  N_\mathrm{events} = \mathcal{T} \int d\lambda ~ P_L (\theta) p_\mathrm{det} (\SNR_c/\SNR),
\end{align}
where $\mathcal{T}$ is the observation time, $p_\mathrm{det}$ is the detection probability of the GW source event \cite{Gerosa:2019dbe}, averaging over the position, inclination, and polarization of the signal, which we obtain using \verb|gwdet| \cite{davide_gerosa_2017_889966}. We use $\SNR_c = 8$ as the threshold for detectable events. The differential element $d \lambda$ which encapsulates the necessary information to estimate source events is given by 
\begin{align}
  d\lambda = \frac{1}{1+z_S} \frac{dR}{dm_1 ~ dm_2} \frac{dV_c}{dz_S} ~ dm_1 ~ dm_2 ~ dz_S.
\end{align}
The comoving volume element
\begin{align}
  \frac{dV_c}{dz} = \frac{4 \pi}{H_0} \frac{r(z)^2}{E(z)}
\end{align}
can be obtained from \verb|astropy| \cite{astropy:2022} for a chosen cosmology, where $r(z)$ is the comoving distance at redshift $z$ and $E(z) = [ \Omega_m(1+z)^3 + \Omega_r(1+z)^4 + \Omega_\Lambda ]^{1/2}$ is the dimensionless Hubble rate. The differential merger rate is given by
\begin{align}
  \frac{dR}{dm_1 ~ dm_2} = &\frac{R_0}{Z_\psi} M^\alpha \eta^\beta \psi(m_1) \psi(m_2) \nonumber \\
  &\times \int dz ~ P_b(z) P_d(t_{\rm delay} (z, z_S)) ,
\end{align}
where $t_{\rm delay}(z, z_S) = t(z_S) - t(z)$ is the delay time between the merger (source) redshift $z_S$ and a formation redshift $z$ (not to be confused with the lensing time delay $t_d$ of \cref{sec:lensing}). The delay time distribution is given by $P_d(t_{\rm delay}) \propto \Theta(t_{\rm delay} - t_{\rm delay, min})/t_{\rm delay}$, where $t_{\rm delay, min}$ is the minimum possible delay time \cite{OShaughnessy:2009szr, Mukherjee:2021qam}. The factor $Z_\psi$ is a normalization constant, such that after integrating over $m_1$ and $m_2$, the local merger rate is given by $R(z=0) = R_0 = 19^{+7}_{-5}$ Gpc$^{-3}$ yr$^{-1}$ \cite{LIGOScientific:2025pvj} following the latest release of LVK data. The quantity $P_b(z)$ is the cosmic star formation rate density, which we take from Ref.~\cite{MadauDickinson} as
\begin{align}
  P_b(z) = 0.015 \frac{(1+z)^{2.7}}{1 + [(1+z)/2.9]^{5.6}} M_\odot \,\text{yr}^{-1}\,\text{Mpc}^{-3}.
\end{align}
For simplicity, the mass distribution of black holes is taken to be
\begin{align}
  \psi(m) \propto m^\zeta.
\end{align}
Finally, we take the mass model used in \cite{Urrutia:2021qak} with $\alpha=0$, $\beta =6$, $\zeta = 0.5$, $m_\mathrm{min} = 3 M_\odot$ and $m_\mathrm{max} = 55 M_\odot$; other black hole mass distributions have been studied in, e.g., Refs.~\cite{Liao:2020hnx, Fairbairn:2022xln, Urrutia:2023mtk}. We have chosen this particular model purely for simplicity, and leave a more careful study of the impact of the BH population modelling for future work, emphasizing that these are only important for forecasting (A+/O5 and ET in this work). Nevertheless, we note that our forecasted constraints for the point lens case are roughly consistent with other works which use different population models~\cite{Jung:2017flg, Liao:2020hnx, Fairbairn:2022xln, Urrutia:2023mtk}.

\begin{figure}
\includegraphics[width=\columnwidth]{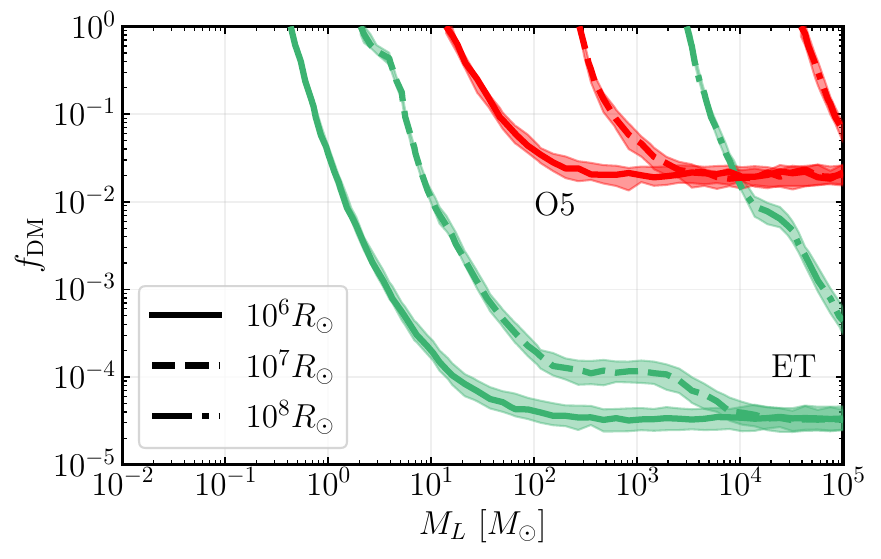}
\caption{\label{fig:f_DM_const_R_projections} Prospective constraints on the extended dark object fraction $f_\mathrm{DM}$ as a function of the compact object mass $M_L$ assuming the non-observation of lensing events at LIGO A+ (red) and ET (green). The compact objects are all of size $R$ and follow a monochromatic distribution. The shaded bands represent regions of uncertainty in the local merger rate, taken to be $R_0 = 19^{+7}_{-5}$ Gpc$^{-3}$ yr$^{-1}$ \cite{LIGOScientific:2025pvj}. }
\end{figure}

\cref{fig:f_DM_const_R_projections} shows the projected reach from LVK O5 (red) and ET (green) as a function of the lens mass for several radii. The central curves within the shaded bands assume the merger rate described above. As expected, the improved sensitivity for O5 results in a factor of $\sim 10$ improvement in the constraint for $f_{\rm DM}$ when compared to current data up to O4. This improvement also results in A+ being sensitive to smaller lens masses by an order of magnitude. On the other hand, we see that ET is expected to provide much stronger reach down to $f_\mathrm{DM} \sim \text{few}\times 10^{-5}$, and extend the sensitivity to sub-solar mass objects. For larger objects, the behaviour of the sensitivity curves is no longer monotonic in the case of ET, as seen from the green curves in \cref{fig:f_DM_const_R_projections}. This is because the increased sensitivity of ET allows for high event SNR, and therefore ET is sensitive to GW lensing events at large $y$. Hence $y_{\rm max}$ falls in the region where the mismatch develops features (near the caustic -- see \cref{fig:y_max_example}), which imprint on the sensitivity curves. In the case of A+ or current LVK data, $y_{\rm max}$ lies below the values of $y$ at which these features in the mismatch appear.

As previously mentioned, in \cref{fig:f_DM_const_R_comparisons} we show the current GW constraints and future projections alongside other existing bounds on constant density extended dark objects, which we obtained from the publicly available repository \cite{Croon:2024jhd}. These bounds are (reading left to right) from the highly magnified extragalactic star Icarus \cite{Croon:2025yfj}, binary evaporation \cite{Ramirez:2022mys}, and the cosmic microwave background \cite{Croon:2024rmw}. Note that the bounds from binary evaporation and the cosmic microwave background were computed for uniform density profiles in the original works, whereas for the lensing of Icarus we extend the procedure in Ref.~\cite{Croon:2025yfj} to uniform density profiles (see \cref{sec:icarus_uniform} for more information). We have also shown constraints on extended dark objects from wide binaries in the ultrafaint dwarf galaxy (UFD) Bo\"otes I \cite{Olea-Romacho:2026pgn}, which use a $\rho \propto r^{-3/2}$ density profile. Constraints from the heating of UFDs \cite{Graham:2023unf, Graham:2025opw} are also shown for comparison, although a detailed study of these bounds for finite-sized objects has not been performed. We have included them because they (along with wide binaries in UFDs) are the most stringent in the mass range probed by O4 and O5, and because they are expected to still apply for sufficiently compact perturbers, similar to what we find for GW lensing. However, the exact physical scale at which they weaken has not yet been studied.

We see that with existing data, the GW lensing constraints are similar to bounds from binary evaporation for small enough objects, but the constraints from UFDs (either wide binaries or dynamical heating) are more stringent than those from GW lensing, even for the future O5 run. 
As expected, all constraints weaken as the lens radius increases (solid versus dashed lines). Crucially, we see that GW lensing probes will continue to improve, eventually providing the leading probe of both point-like and extended lenses. 
Furthermore, due to the increased sensitivity of these detectors, they will be more sensitive to larger (and therefore, more diffuse) lenses.
With O5 scheduled for 2027, we expect an order-of-magnitude increase in the sensitivity to $f_{\rm DM}$ from available LVK data within the next two years~\cite{Basak:2021ten}.

\section{Constraints on Microphysics} \label{sec:compact_objects}
In this section we translate GW lensing bounds and projections into the parameter space of two benchmark models that produce extended dark objects: dissipative dark sectors (\cref{sec:adm}) and light scalar fields that form boson stars (\cref{sec:boson_stars}). We also present constraints and projections on the DM fraction in extended objects in the mass-density or mass-radius parameter space in \cref{sec:general_constraints}; these can be easily translated into microphysical parameters in other scenarios. 

\subsection{Dark stars from dissipative DM}
\label{sec:adm}
\begin{figure*}
\includegraphics[width=\textwidth]{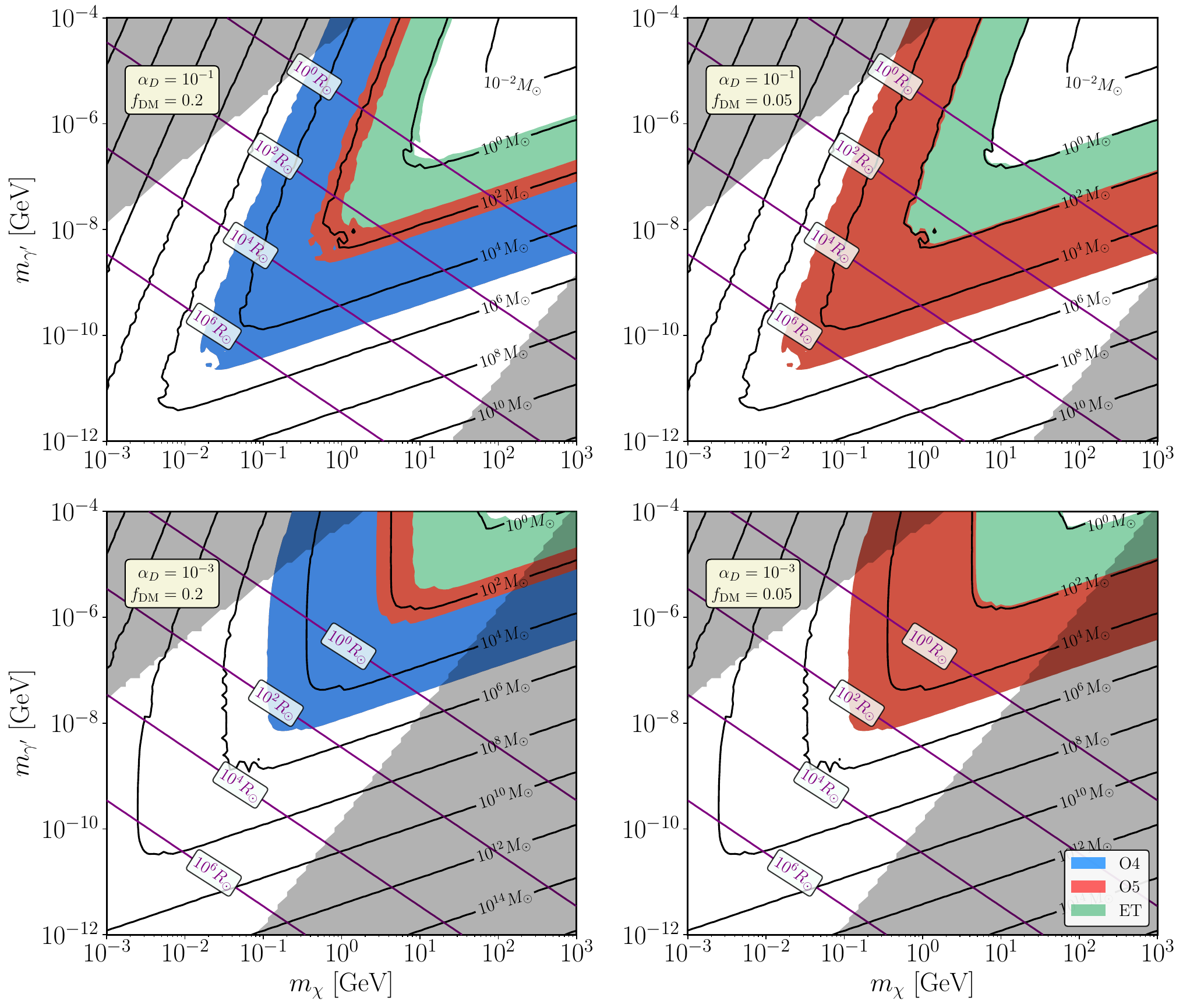}
\caption{\label{fig:dco_constraints} Current and prospective constraints from diffractive GW lensing on the parameter space for the dark electron and dark photon model, assuming that the dark asymmetric stars follow a monochromatic distribution with $f_\mathrm{DM}= 0.2$ in the left panels and $f_{\rm DM} = 0.05$ in the right panels. The blue region shows current constraints from available GW data up to GWTC-5/O4, whereas the red and green regions are prospective constraints from A+ and ET, respectively. The grey shaded region is where one expects black holes. Solid black contours are contours of constant dark object masses. Solid purple lines are contours of constant dark object radii. For all of the parameter space, the asymmetric dark stars are sufficiently dense that the lensing effects are well in the point-like regime. Note that the O4 and O5 regions lie on top of the ET region. We have assumed $\alpha_D = 10^{-1}$ and $\alpha_D = 10^{-3}$ in the top and bottom panels, respectively.}
\end{figure*}

Dissipative DM models are a natural extension of the minimal elastic scattering-only models of self-interacting DM (SIDM)~\cite{Tulin:2017ara}. By introducing inelastic scattering, a gas of DM particles is not only able to re-distribute energy throughout itself, but it can also cool and dissipate away its kinetic energy through the emission of a light mediator. These models often have significant cosmological and astrophysical implications, particularly for very small-scale structure, resulting in the formation of stable, self-gravitating, extended dark objects on the scales of stars and planets. See \cite{Cline:2021itd} for a review. Such dissipative dark sectors can make up a significant component of the matter budget of the universe (up to $\sim 10\%$ of DM) without running afoul of SIDM or galactic morphology constraints \cite{Fan:2013tia, Fan:2013yva,Schutz:2017tfp, Bansal:2022qbi, Barron:2025dys}. 

We consider a minimal dark sector composed of a dark Dirac fermion $\chi$ (the dark electron) with mass $m_{\chi}$ and a massive dark U(1) gauge boson $A^\mu$ (the dark photon) with mass $m_{\gamma'}$. The Lagrangian can be written as 
\begin{equation}
  \mathcal{L} \supset \bar{\chi} (i \gamma^\mu D_\mu - m_\chi) \chi - \frac{1}{4} F_{\mu \nu} F^{\mu \nu} + \frac{1}{2}m_{\gamma'}^2 A_\mu A^\mu,
  \label{eq:dissipative_model}
\end{equation}
where $D_\mu = \partial_\mu - i g_D A_\mu$ is the gauge covariant derivative with the coupling parameter $\alpha_D = g_D^2/(4\pi)$, and $F_{\mu \nu} = \partial_\mu A_\nu - \partial_\nu A_\mu$ is the field strength tensor. Additionally, we assume that there is an asymmetry between particles and anti-particles (i.e., dark positrons $\bar{\chi}$ are absent) following~\cite{Chang:2018bgx}; this choice avoids complications associated with bound state formation and $\chi\bar{\chi}$ annihilation during structure formation. 

Linear and non-linear structure formation in these models has been studied semi-analytically in~\cite{Chang:2018bgx, Bramante:2023ddr, Bramante:2024pyc}. Dark electron density perturbations on galactic scales track those of cold dark matter (CDM) during linear evolution as long as the mass of the perturbation $M$ is larger than the Jeans mass~\cite{Chang:2018bgx}
\begin{align}
  m_J = \frac{\pi}{6} \left( \frac{\pi}{G} \right)^{3/2} c_s^{3} \left( \frac{1}{m_\chi n_\chi} \right)^{1/2},
  \label{eq:jeans_mass}
\end{align}
where 
\begin{align}
  c_s = \sqrt{\frac{T_\chi}{m_\chi} + \frac{4 \pi \alpha_D n_\chi }{m_\chi m_{\gamma'}^2} }
  \label{eq:soundspeed}
\end{align}
is the sound speed of the dark electron gas. 
 The first term in $c_s$ is due to the kinetic pressure of the gas and the second term is due to the repulsive dark-photon-mediated interactions~\cite{Kouvaris:2015rea}.
The perturbations eventually decouple from the Hubble flow and turn around. The subsequent non-linear dynamics can be qualitatively understood by considering the dark electron density $n_\chi$ and temperature $T_\chi$ as the halo begins to contract. 
 The halo first adiabatically free-falls, collapsing on a free-fall timescale $t_{\rm collapse} \sim t_{\mathrm{ff}} = (16\pi \rho G)^{-1/2}$, since $M > m_J$ and cooling is still irrelevant due to the low temperature and density. The free-fall compresses the dark electron gas, leading to an increase in the kinetic pressure and consequently the Jeans mass. Once the Jeans mass is comparable to the mass of the halo, $M\sim m_J$, 
further contraction is only possible due to cooling of the dark electrons via $\chi$--$\gamma'$ interactions. This leads to a cooling-limited contraction phase (called ``nearly-virialized contraction'') with a characteristic timescale $t_{\rm collapse} \sim t_{\rm cool}$, where $ t_{\rm cool}$ is the timescale associated with the gas cooling reactions available in the dark sector. In the model of \cref{eq:dissipative_model}, the only relevant cooling channel is bremsstrahlung $\chi \chi \to \chi\chi+ \gamma'$. 
Eventually cooling becomes so efficient that $t_{\rm cool} \sim t_{\rm ff}$ and the gas collapses on gravitational timescales again. Moreover, as this happens the Jeans mass rapidly shrinks below the halo mass, which enables different regions within the halo to undergo their own collapse. This leads to fragmentation: the halo divides itself into smaller clumps, which can each continue to collapse. The main features of the evolution of dark electron halos are similar to the initial stages of star formation~\cite{Bromm:2013iya}; the resulting dark electron fragments are the dark analogues of the protostars in the visible sector.

A key result in Ref.~\cite{Chang:2018bgx} was that in this simple dark sector, the final fragment masses and sizes are independent of the collapse trajectory prior to fragmentation, as long as fragmentation begins. This means that for a given set of microphysical parameters, the smallest (minimal) fragment size should be the same in different galaxies.
The properties of the minimal fragments are determined by fragmentation termination, which occurs either because the dark electron halo becomes optically thick to emitted dark photons or because the pressure due to $\chi$--$\chi$ self-interactions (i.e., the second term in \cref{eq:soundspeed}) halts the collapse. The Jeans mass at the end of fragmentation gives the final mass of the fragments. If the fragment is opaque to dark photons, we assume that the star can cool via surface cooling and further collapse until it is pressure-supported by the $\chi$--$\chi$ repulsion. Hence, when the kinetic term is subdominant in \cref{eq:soundspeed}, the Jeans mass is no longer dependent on temperature. The final radii of the dark stars are then given by $m_J = 4\pi m_\chi n_\chi R^3/3$, which gives
\begin{align}
  R = \frac{\pi}{m_\chi m_{\gamma'}} \sqrt{\frac{\alpha_D}{G}}. 
\end{align}
The mass of these objects depends on the detailed evolution of the halo through the $T_\chi$--$n_\chi$ space as described in \cite{Chang:2018bgx}, and is also determined by $\alpha_D$, $m_\chi$ and $m_{\gamma'}$. The evolution is obtained by solving a simple thermodynamical system \cite{Chang:2018bgx} 
\begin{align}
  \frac{d \log T_\chi}{d\log\rho_\chi} = \frac{2}{3} \frac{m_\chi P_\chi}{\rho_\chi T_\chi} - 2 \frac{t_{\rm collapse}}{t_{\rm cool}},
\end{align}
until one of the fragmentation conditions previously discussed is reached, where $P_\chi$ is the pressure of the dark electron gas. Remarkably, the minimal sizes and masses of the asymmetric dark stars are fully specified by the microphysical model. We note that the cosmological or particle physics details of a given model can modify the formation history and final properties of the dark stars~\cite{Chang:2018bgx, Bramante:2023ddr,Bramante:2024pyc}.

In order to study the lensing properties of dark stars, we need to model their density profiles. Refs.~\cite{Kouvaris:2015rea,BetancourtKamenetskaia:2026cyf} solved the Tolman-Oppenheimer-Volkoff equations for these objects, finding constant density cores up to nearly the star radius, followed by a very sharp drop. In other words, a uniform density profile is a reasonable approximation for these numerical solutions.

In \cref{fig:dco_constraints} we show the predictions for the final fragment masses in the $m_\chi$--$m_{\gamma'}$ plane. The black (purple) curves indicate contours of constant final fragment masses (radii).\footnote{The small irregularities in the contours are due to the finite grid on which the fragment masses were computed -- see \cite{Bramante:2024pyc}.} We have chosen two different values of the coupling, with $\alpha_D = 10^{-1}$ in the top panels and $\alpha_D = 10^{-3}$ in the bottom panels. The regions shaded in grey show where one would expect a black hole to form, which we estimate by comparing the physical radius of the object with its Schwarzschild radius $2 GM$. We also show current constraints from LVK O4 in solid blue, along with projected sensitivity from A+ and ET as the solid red and green regions, respectively. 
We have truncated the constraints at $M_L = 10^{5} M_\odot$ to avoid the strong lensing regime with multiple images.
If 20\%\footnote{Although it is difficult to have 20\% of DM be dissipative (see e.g. \cite{Fan:2013tia, Fan:2013yva, Schutz:2017tfp, Bansal:2022qbi, Barron:2025dys} for the case of atomic dark matter), we show these constraints to demonstrate the current reach of LVK O4.} of the DM is in these objects, the currently available LVK data is able to constrain a large region of this parameter space, ruling out fragment masses in the range $\sim10^2 M_\odot$ to $10^5 M_\odot$. Future data from LVK O5 (ET) will be able to probe the existence of dark stars down to $M\sim 10 M_\odot$ ($M\sim 1 M_\odot$) with abundances down to $\text{few}\times 10^{-2}$ ($\text{few}\times 10^{-5}$ -- see \cref{fig:f_DM_const_R_comparisons}), which corresponds to higher dark electron and dark photon masses. These bounds complement the projected sensitivity of future observatories derived from GWs from the mergers of dark stars~\cite{BetancourtKamenetskaia:2026cyf}. 
Interestingly, we find that the dark object densities predicted in this scenario are sufficiently high that a point lens provides an adequate description of the physics where GW observatories have sensitivity. Thus PBH constraints from GW diffraction can be directly translated onto this parameter space. This model therefore provides a minimal example in which microphysical parameters (particle masses and couplings) can be mapped to macroscopic properties (dark star sizes and masses), which are constrained by observations. 
It would be interesting to see how GW diffraction can be used to constrain other models of dissipative dark sectors, such as atomic dark matter~\cite{Kaplan:2009de}. 

\subsection{ALP stars from post-inflationary Peccei-Quinn breaking}
\label{sec:boson_stars}

\begin{figure*}
\includegraphics[width=\textwidth]{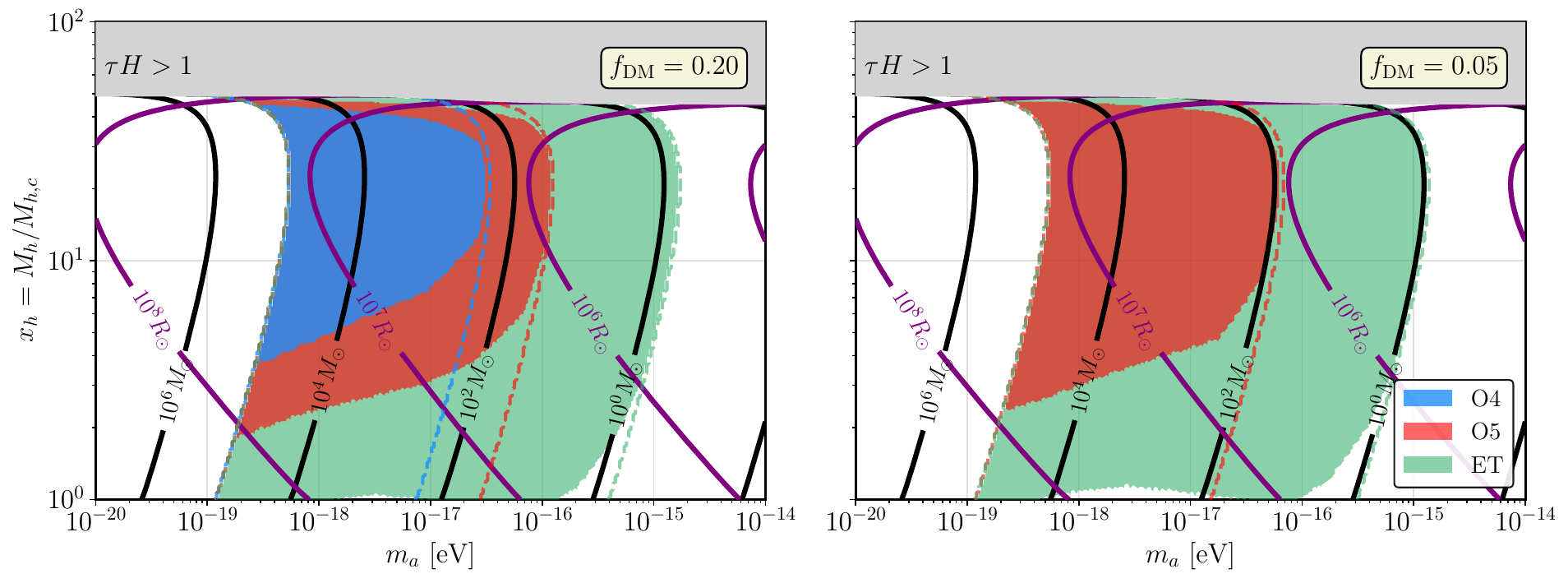}
\caption{\label{fig:axion_star_constraints} GW lensing constraints and projections for axion-like particle (ALP) stars produced from post-inflationary Peccei-Quinn symmetry breaking \cite{Chang:2024fol} in the ALP mass - minihalo mass parameter space. The minihalo mass is given as a fraction $x_h$ of the characteristic mass in \cref{eq:characteristic_minihalomass}. The black (purple) contours represent constant mass (radius) of the ALP stars. The region in grey indicated by $\tau H > 1$ is where the ALP stars are unable to form as the condensation timescales are longer than the Hubble time at matter-radiation equality when the first miniclusters form. We have shown current constraints from O4 in the blue shaded region, along with prospective constraints from A+ and ET in the red and green shaded regions, respectively. We have assumed that the ALP stars follow a monochromatic distribution, and make up $f_{\rm DM} = 0.20$ and $0.05$ of the DM in the left and right panels, respectively. The parameter space between the dashed lines represents constraints assuming point lenses for each detector.}
\end{figure*}

Axions and axion-like particles (ALPs) are a class of bosonic DM candidates which can form small-scale structure. 
Axions are particularly well-motivated as a potential solution for the strong charge-parity problem in quantum chromodynamics (QCD) \cite{PecceiQuinn}. The cosmological and astrophysical implications of the QCD axion and ALPs have been studied extensively in the literature (see e.g. Refs. \cite{Marsh:2015xka, Ferreira:2020fam, Hui:2021tkt, Chadha-Day:2021szb} for reviews). Minimal QCD axion models make a concrete prediction for the axion mass as a function of its coupling to SM particles, while in ALP models the couplings and masses are uncorrelated; we use the terms ``axion'' and ``ALP'' interchangeably below and remain agnostic about the origin of its mass. As DM candidates, ALPs are much lighter than particle DM such as that considered in \cref{sec:adm}, with masses $\ll 1$ eV and as light as $3\times 10^{-19}$ eV~\cite{May:2025ppj} (even lower-mass fields can affect cosmological observables and are often called fuzzy DM~\cite{Hu:2000ke, Hui:2016ltb}). This means that ALPs must be produced non-thermally in the early universe and, as we describe below, typical production mechanisms endow ALPs with small-scale isocurvature fluctuations that enable them to collapse into dense structures well before galactic-scale structure formation begins. 

Axions and ALPs can be produced either before or after inflation through several mechanisms, including misalignment: the displacement of the ALP field from the minimum of its potential. 
If misalignment occurs after inflation, different causally-disconnected regions feature different misalignment field values, and the resulting patchwork of field domains corresponds to isocurvature fluctuations in the DM density field. 
Once oscillations of the field commence at $m_a \sim H(T_{\rm osc})$, its time-averaged dynamics on sufficiently large scales are identical to those of CDM; on smaller scales (dictated by the ALP mass) significant differences emerge due to the non-negligible field gradient effects. 
The relative density fluctuations are large and lead to the formation of axion miniclusters shortly after matter-radiation equality~\cite{Kolb:1993hw}. These gravitationally-bound objects are much more compact than the smallest halos expected in CDM with a Harrison-Zeldovich initial spectrum by virtue of their early formation time (recall that the typical density of a virialized halo can be estimated as $\sim 200\bar{\rho}(z_\mathrm{collapse})$). However, these early miniclusters are not dense enough to induce significant GW lensing. Detailed studies of the internal dynamics of the miniclusters reveal that their centres generally form solitons called axion stars.\footnote{Similar to the dark stars of \cref{sec:adm}, ``axion stars'' is something of a misnomer since there is no fusion -- these objects are supported by wave effects or self-interactions.} These substructures are stabilized by the competition of gravity and field gradient effects (as opposed to gravity and kinetic pressure for miniclusters)~\cite{Visinelli:2017ooc,Chang:2024fol} (and field self-interactions in certain cases~\cite{Kolb:1993hw,Chavanis:2011zi}). More precisely, minimizing the total (gravitational plus field gradient) energy of an axion star with mass $M_\star$ with respect to its radius $R_\star$ yields~\cite{Visinelli:2017ooc,Chang:2024fol}
\begin{align}
  R_\star \approx 1.7 \times 10^7 R_\odot \left( \frac{10^{-18} \text{ eV}}{m_a} \right)^2 \left( \frac{2\times 10^4 M_\odot}{M_\star} \right).
  \label{eq:axion_star_mass_radius_relation}
\end{align}
Comparing this with the sensitivity in \cref{fig:f_DM_const_R_comparisons} suggests that axion stars \emph{are} suitable targets for GW diffraction searches.
Additionally, $M_\star$ is not a free parameter, but depends on the axion mass and host minicluster mass as we discuss below; in \cref{eq:axion_star_mass_radius_relation} $M_\star$ is normalized to a characteristic value expected at that $m_a$.

Axion production prior to inflation is not typically associated with 
enhanced DM substructure, since inflation ensures that the initial misalignment is the same across all of the Hubble patches that eventually make up the observable universe. However, this neglects non-linear axion dynamics which can arise in models with non-quadratic ALP potentials. In particular, large values of the initial misalignment can result in rapid growth of axion density fluctuations on certain scales~\cite{Arvanitaki:2019rax}, which also leads to the formation of miniclusters and solitons. Therefore, both pre- and post-inflationary axion production can result in a population of axion stars. 

To demonstrate the potential sensitivity for GW diffraction effects from axion substructure, we follow Ref.~\cite{Chang:2024fol} in outlining the connection between the axion, minicluster and star masses relevant for GW lensing. First, we note that the characteristic mass of a density perturbation that collapses into miniclusters is determined by the scale at which the axion density power spectrum peaks, which is of order the horizon size $k_{\rm osc}^{-1}$ at the time when axion oscillations begin; the mass contained within this horizon is 
\begin{align}
  M_{h,c} \sim \frac{4\pi}{3} \left( \frac{1}{k_{\rm osc}} \right)^3 \bar{\rho}_{a,0},
  \label{eq:characteristic_minicluster_mass_generic}
\end{align}
where $k_{\rm osc} = a_{\rm osc} H(T_{\rm osc}) \sim k_{\rm eq} \sqrt{m_a/H_{\rm eq}}$ and $k_{\rm eq} = a_{\rm eq }H_{\rm eq}$ is the comoving wavenumber corresponding to matter-radiation equality ($H_{\rm eq} \approx 2.2\times 10^{-28}$ eV). The precise proportionality coefficient depends on the axion model (QCD axion versus axion-like particle or pre- or post-inflationary misalignment). For ALPs produced after inflation, \cref{eq:characteristic_minicluster_mass_generic} evaluates to~\cite{Chang:2024fol}
\begin{equation}
  M_{h,c} \approx 3.1 M_\odot \eta \left( \frac{10^{-16} \text{ eV}}{m_a} \right)^{3/2},
  \label{eq:characteristic_minihalomass}
\end{equation}
where $\eta = g_{*S}(T_{\rm osc})/g_*^{3/4}(T_{\rm osc})$ varies between $1.6$ and $1.8$ for $m_a \in [10^{-19}, 10^{-14}]$ eV.

Realistic power spectra lead to the formation of a distribution of minicluster masses $M_h$, which can be estimated using simulations or via the semi-analytic Press-Schechter formalism. Axion star formation depends on the local conditions within a particular minicluster, and therefore on $M_h$; following \cite{Chang:2024fol} 
we treat $x_h = M_h/M_{h,c}$ as a free parameter without imposing a particular mass function. 

Axion stars form within a minicluster when virialized axions Bose-condense into a lower energy state via scattering; this scattering is mediated by gravity or self-interaction and is enhanced by the enormous axion occupation numbers~\cite{Dmitriev:2023ipv}. The final axion star mass $M_\star = M_h x_\star(m_a, M_h)$ is determined by the minicluster mass fraction $x_\star$, which was computed in~\cite{Dmitriev:2023ipv} for an isolated minicluster. Following \cite{Chang:2024fol} we evaluate $x_\star$ at matter-radiation equality, which is appropriate for axion stars formed in the first generation of miniclusters. 
We find in this way that $x_\star$ can be tens of percent for $x_h$ of order a few (consistent with Fig. 3 of \cite{Chang:2024fol}). 
While the axion stars continue to grow after equality, minicluster mergers can disrupt this growth, or lead to the merger of the stars themselves. Thus, this estimate corresponds to the least massive gradient-supported objects that can form.

The Poisson-Schr\"odinger equation that determines the balance between gravity and field gradient pressure can be solved numerically to determine the density profile of a boson star. The resulting profiles can be approximated by \cite{Schive:2014dra}
\begin{equation}
  \rho(r) \propto \frac{1}{(1 + 0.091 (r/r_c)^2)^8},
  \label{eq:alp_star_density_profile}
\end{equation}
which is valid for $r \lesssim 3r_c$, with an NFW profile providing a better description at larger radii in simulations. The NFW part of the profile corresponds to the virialized minicluster ALPs, rather than the solitonic core. We therefore identify $R_\star$ from \cref{eq:axion_star_mass_radius_relation} with the approximate transition radius at $3r_c$. The steep fall-off of \cref{eq:alp_star_density_profile} beyond $3r_c$ suggests that our uniform ball lens would be a reasonable approximation for this density profile. 
For ALPs with non-quadratic potentials (e.g., the QCD axion) self-interactions can be important in determining the density profile; while the precise form differs from the minimal ALP model, see e.g. \cite{Eby:2018dat}, these density profiles all feature a core and a sharp fall-off beyond some characteristic radius, which again suggests the utility of the constant density ball as a toy model for these objects.

\cref{fig:axion_star_constraints} shows the parameter space of ALP stars as a function of $m_a$ and $x_h = M_h/M_{h,c}$, with the $R_\star(m_a, x_h)$ and $M_\star(m_a, x_h)$ contours computed using the precise expressions collected in \cite{Chang:2024fol}. We find that the soliton masses can span $10^{-12} M_\odot$ to $10^8 M_\odot$ for $m_a$ between $10^{-20}$ eV and $10^{-6}$ eV. The GW diffraction sensitivity of current and future detectors is shown by the shaded regions as before. 
Unlike the dark stars of the previous section, the compactness of the solitons can vary widely, so that their extended nature has a significant impact on the GW lensing. This can be seen by comparing the solid regions to the corresponding dashed lines which assume a point-like lens.

\subsection{General constraints on lens mass and density}
\label{sec:general_constraints}
\begin{figure}
\includegraphics[width=\columnwidth]{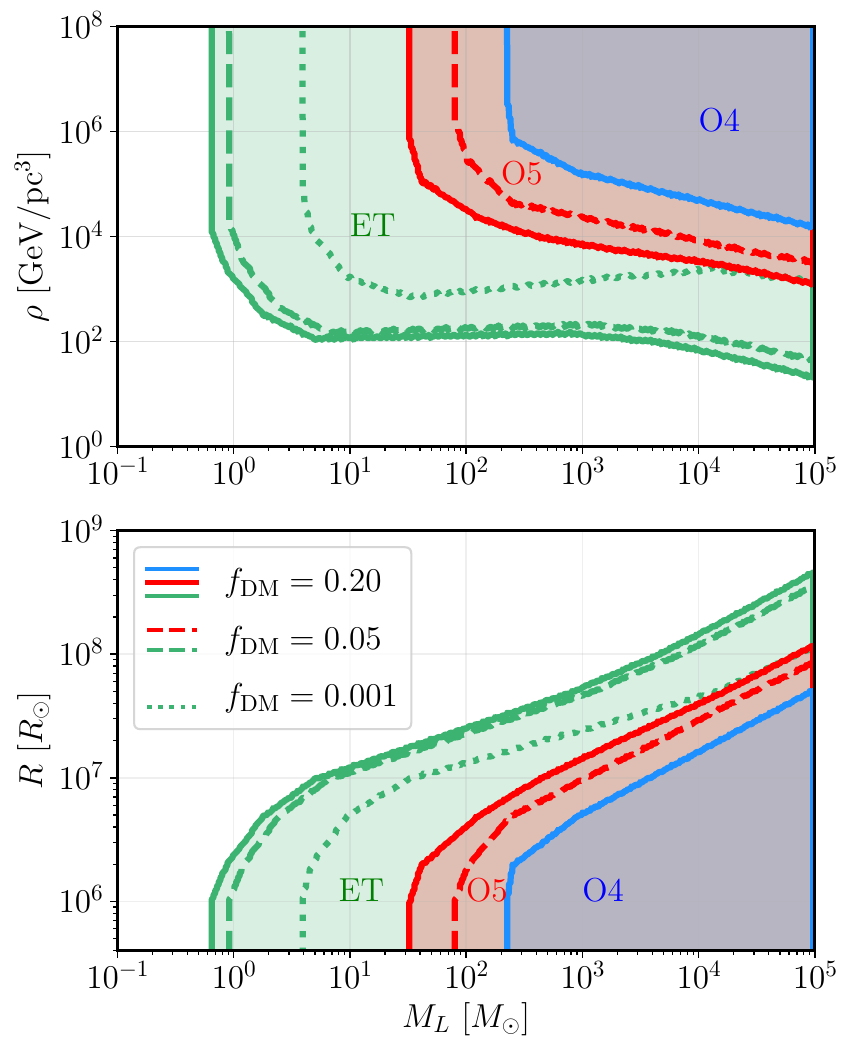}
\caption{\label{fig:mass_vs_dens_and_R} Current and prospective general constraints in the $M_L$-$\rho$ ($M_L$-$R$) parameter space in the top (bottom) panel for the uniform ball lens from the non-observation of GW diffraction. The shaded contours represent constraints from LVK O4 (blue), as well as from the upcoming LVK O5 (red) and the Einstein Telescope (green). We assume $f_{\rm DM} = 0.20, 0.05,$ and $0.001$ in solid, dashed, and dotted curves, respectively.}
\end{figure}
We end this section by showing in \cref{fig:mass_vs_dens_and_R} the constraints on the lens mass-density parameter space assuming uniform ball lenses, for three different extended object fractions $f_{\rm DM} = 0.20, 0.05,$ and $0.001$ in the solid, dashed, and dotted lines, respectively. Note that these sensitivity levels are not reachable by all of the detectors. LVK O4 is only sensitive to $f_{\rm DM} = 0.2$ and LVK O5 will be sensitive to $f_{\rm DM} = 0.2$ and $0.05$, whereas ET will be sensitive to all of these levels. These constraints on the mass and density of the extended lenses are general and can thus be applied to any underlying model, such as dissipative models with various cooling channels~\cite{Bramante:2023ddr}, primordial neutron stars~\cite{Krnjaic:2026sgl} and non-ALP bosonic objects~\cite{Eby:2015hsq}. 

\section{Conclusion} \label{sec:conclusion}

In this work, we have demonstrated how the non-point-like nature of extended dark matter objects modifies diffractive gravitational lensing of GWs compared to black holes. In general, constraints on the fraction of DM in point-like lenses cannot be simply applied to extended lenses if they are larger than their Einstein radius. Using a uniform density ball as a toy model for these objects, we re-interpreted the non-observation of lensing in GWTC-5 as a constraint on the fraction of DM for a variety of lens masses and radii. We expect that our findings are qualitatively general in that the sensitivity to these extended lenses will generally weaken as the size of the object increases due to the flattening of the gravitational potential.

We have also shown how such bounds can be mapped onto the parameter space of the underlying microphysical model that produces the dark extended lenses in the context of two examples: dissipative dark sectors and axion-like particles. Thus GW diffraction provides a viable method for testing interesting dark sector dynamics even in the absence of non-gravitational couplings to the Standard Model particles. 

There are several follow-up studies that would complement the work in this paper. First, it would be straightforward to re-interpret the GW diffraction constraints in any other models that predict extended dark matter objects (such as those mentioned in \cref{sec:general_constraints}), following our two examples. 
Our analysis can also be improved in a few different ways.
Throughout this work we assumed constant density profiles and monochromatic mass distributions. Both of these details are in principle calculable for a given microphysical dark sector model and its cosmological evolution. While this may be difficult in practice (e.g. the prediction of the dark star mass function in dissipative models is analogous to computing the standard initial mass function from first principles), these are important inputs for deriving precise constraints. We have also made several approximations in our statistical treatment of the detectability of lensed events and the resulting constraints on the DM fraction. While these approximations are valid in the high SNR limit for certain choices of Bayesian priors, a clear (but computationally expensive) improvement is to perform a full Bayesian model comparison following the LVK lensing analyses. Another important assumption is that the SNR of lensed events as would be observed by, e.g., LVK, is close to the optimal value $\ip{\hL}{\hL}$; as shown in \cite{Chan:2024qmb} this is not always the case if the search is performed using an unlensed template bank and the wave effects are strong (e.g. for small impact parameters and large lens masses). A signal injection study is then needed to evaluate the true SNR for this matched-filter search. 

Finally, in this work we have not discussed the presence of ``background'' lensing events due to astrophysical lenses such as black holes. Future detectors like LVK O5 and ET, however, will very likely observe GW lensing.\footnote{Note that already in GWTC-4.0, one possible lensing candidate has been announced by the LVK Collaboration, which requires further study \cite{LIGOScientific:2025cwb}; we have not included it in our analysis. However, if this candidate is confirmed as a genuine lensing event, the main results of our work will remain qualitatively the same. We would simply need to increase the lensing threshold from 3 to 3.9 events when setting bounds on the fraction of DM in lenses, see e.g. \cite{Croon:2020ouk}.} In this case, it will be important to account for proper modelling of the background events in order to place constraints on the population of extended dark objects, and consequently the particles that make them up. We leave this for future study.

\begin{acknowledgments}
We thank Reed Essick, Paddy Fox, Jose Mar\'ia Ezquiaga, and Sergio Sevillano Mu\~noz for useful discussions.
This work was supported by the Natural Sciences and Engineering Research Council of Canada (NSERC). Our numerical calculations were enabled in part by support provided by Compute Ontario and the Digital Research Alliance of Canada (\href{http://alliancecan.ca}{alliancecan.ca}).
\end{acknowledgments}

\onecolumngrid
\appendix
\setcounter{figure}{0}
\renewcommand{\thefigure}{A\arabic{figure}}

\crefalias{section}{appendix}
\section{Connection to Bayesian Analysis}
\label{sec:stats}
In this Appendix we outline the approximations leading to the semi-analytic statistical treatment used in the main text. An alternative to this approach is a numerical calculation of the relevant Bayes factors, which is computationally intensive, requiring integrations over multidimensional posteriors for lensed and unlensed waveform models across a large parameter space. We will follow \cite{Cornish:2011ys,Vallisneri:2012qq,DelPozzo:2014cla} to argue that the two approaches should yield similar results for high SNR events.

We want to approximate the Bayes factor
\begin{equation}
 \mathcal{B}\;=\; \frac{P(H_1 \,|\, d)}{P(H_0 \,|\, d)}
 \;=\; \frac{\int d\theta\, d\lambda\;
    p(d \,|\, \theta, \lambda)\, \pi(\theta, \lambda)}
    {\int d\theta\;
    p(d \,|\, \theta)\, \pi(\theta)} ,
 \label{eq:bayes_factor_def}
\end{equation}
where $P(H_1|d)$ is the probability that the lensed hypothesis is true
given the data, and $P(H_0|d)$ the same for the unlensed hypothesis; the
first equality assumes equal prior odds for the two hypotheses. Here
\begin{equation}
 \lambda = \{ M_L, R, y \}, \qquad
 \theta = \{ \text{GW parameters} \},
\end{equation}
$p(d|\cdots)$ are likelihoods and $\pi(\cdots)$ are prior probabilities.
We write $\Dth = \dim\theta$ and $\DL = \dim\lambda$ (here $\DL = 3$).

The likelihoods are Gaussian,
\begin{equation}
 p(d \,|\, \theta, \lambda)
 = \mathcal{N}\,
  e^{-\ip{d - h(\theta,\lambda)}{d - h(\theta,\lambda)}/2},
 \label{eq:gaussian_likelihood}
\end{equation}
where the noise $n$ satisfies
$\big\langle \ip{a}{n}\ip{n}{b} \big\rangle = \ip{a}{b}$ and $\mathcal{N}$ is a normalization constant. We will denote the lensed waveform as $h(\theta,\lambda)$, and 
the unlensed one as $h(\theta)$. First, assume the data contains lensing:
\begin{equation}
 d = h(\theta_\true, \lambda_\true) + n .
\end{equation}
For sufficiently strong signals, we expect the likelihood to be narrowly peaked, meaning 
most of its support comes from a small region of parameter space near the true values, enabling the Laplace approximation~\cite{Kass:1995loi}.

Expanding around the maximum likelihood (ML) point, the Gaussian integrals for the lensed evidence give~\cite{Vallisneri:2012qq}
\begin{equation}
 \int p(d \,|\, \theta, \lambda)\, \pi(\theta, \lambda)\,
   d\theta\, d\lambda = \mathcal{N}\,
  e^{-\ip{n}{n}/2
    + \frac{1}{2}(G^{-1})^{\mu\nu}
     \ip{n}{\partial_\mu h}\ip{n}{\partial_\nu h}} 
  \times\pi(\theta_\ML, \lambda_\ML)\,
  (2\pi)^{(\Dth + \DL)/2} \sqrt{|G^{-1}|},
 \label{eq:lensed_evidence}
\end{equation}
where the prior has been pulled out of the integral (assuming it is flat over the support of the likelihood); here the Greek letters index both $\lambda$ and $\theta$ parameters and $G = G_{\alpha\beta} \equiv \ip{\partial_\alpha h}{\partial_\beta h}$ is the Fisher matrix of the lensed model. The factor $\sqrt{|G^{-1}|}$ is (proportional to) the posterior
uncertainty volume; $\Dth + \DL$ is the total dimensionality of the Gaussian integral. Note that the validity of this formula requires that the Fisher matrix is invertible. At small true lens masses, or large impact parameters, the lens parameters become non-identifiable (i.e., the likelihood is flat along these directions), so a Gaussian approximation is not valid along every direction in $\lambda$. 

Repeating the same approximations for the unlensed model evidence (the denominator in \cref{eq:bayes_factor_def}) gives a similar expression to \cref{eq:lensed_evidence}, but with $\pi(\theta, \lambda) \to \pi(\theta)$, $(\Dth + \DL) \to \Dth$ and $G\to F$, with $F$ being the (smaller) Fisher matrix of the unlensed hypothesis. Taking the ratio of the two evidences yields the approximate Bayes factor:
\begin{align}
 \ln \mathcal{B} & = \tfrac{1}{2}\ip{\dhp}{\dhp} + \ip{\dhp}{n} + \tfrac{1}{2}\Big[ \left(G^{-1}\right)^{\mu\nu}
    \ip{n}{\partial_\mu h}\ip{n}{\partial_\nu h}
   - \left(F^{-1}\right)^{ij}
    \ip{n}{\partial_i h}\ip{n}{\partial_j h} \Big] + \frac{1}{2}c_L,
 \label{eq:lnB_lensed}
\end{align}
with
\begin{equation}
  c_L = \DL \ln 2\pi + 2\ln \frac{\pi(\theta_\ML, \lambda_\ML)}{\pi(\theta_\ML)}\frac{\sqrt{|G^{-1}|}}{\sqrt{|F^{-1}|}}
  \label{eq:lensed_occam}
\end{equation}
and where Ref.~\cite{Vallisneri:2012qq} defines 
\begin{equation}
 \dhp \;\equiv\; \delta h
  - \left(F^{-1}\right)^{ij} \partial_j h \ip{\delta h}{\partial_i h(\theta)}
\end{equation}
with $\delta h = h(\theta_\ML, \lambda_\ML) - h(\theta_\ML)$. $\dhp$ has a nice interpretation -- it is the difference between lensed and unlensed waveforms that is orthogonal (in the noise-weighted inner product sense) to the space of unlensed waveforms; i.e. the unlensed waveform parameters cannot be adjusted to absorb this difference.
Ref.~\cite{Vallisneri:2012qq} shows that $\ip{\dhp}{\dhp}$ is equivalent to the minimized mismatch in \cref{eq:non_centrality,eq:mismatch} to $\mathcal{O}(1/\SNR^4)$. 
 
$c_L$ in \cref{eq:lensed_occam} is the Occam factor; it penalizes models for having extra parameters and measures how much the posterior volume is reduced compared to the prior volume~\cite{Cornish:2011ys}. 

Note that \eqref{eq:lnB_lensed} is \emph{quadratic}
in the noise which suggests that $2\ln \mathcal{B}$ is
$\chi^2$-distributed. We can match moments to figure out what the relevant mean and number of degrees of freedom are. In particular, it is straightforward to compute the mean and variance 
\begin{align}
  \langle \lnB - c_L/2\rangle & = \tfrac{1}{2}\ip{\dhp}{\dhp} + \DL/2 \\
  \big\langle (\lnB - c_L/2)^2 \big\rangle - \langle \lnB -c_L/2\rangle^2
 & = \ip{\dhp}{\dhp} + \DL/2 ,
\end{align}
which imply \cref{eq:lnB_under_HL}, i.e. $2\lnB-c_L$ is distributed according to a non-central $\chi^2$ distribution with $D_\lambda$ degrees of freedom and non-centrality parameter $\ip{\dhp}{\dhp}$. 

The above calculations of $\lnB$ can be repeated assuming the data is unlensed; we find \cref{eq:lnB_under_null}, i.e. $2\lnB-c_U$ is distributed according to the central $\chi^2$ distribution with one degree of freedom ($c_U$ is the Occam factor for this scenario). The non-centrality parameter vanishes because the unlensed waveforms are a subset of the lensed model, so at the ML point $\dhp \to 0$. The single degree of freedom requires a little elaboration. We expect the ML parameter point of the lensed hypothesis to be pushed towards small $M_L$ where the other lensing parameters $R$ and $y$ have no effect -- they become flat directions. Evidence integrals along these directions are therefore prior (rather than likelihood) dominated. We assume that the likelihood in the remaining $M_L$ direction is well approximated by a Gaussian, which gives the single degree of freedom. There is an additional complication in that $M_L = 0$ is a physical boundary of the parameter space -- see \cref{fn:stat}. Ref.~\cite{Algeri:2019lah} describes how this modifies the distribution; we do not pursue this improvement here. 

Knowing the distributions of $\lnB$ for lensed and unlensed data allows us to relate the lensing threshold $\Bthr$ to the false positive probability as described in \cref{sec:det_crit}.
\section{Mismatch Minimization}
\label{sec:mm_minimization}

\begin{figure}
\includegraphics[width=0.95\columnwidth]{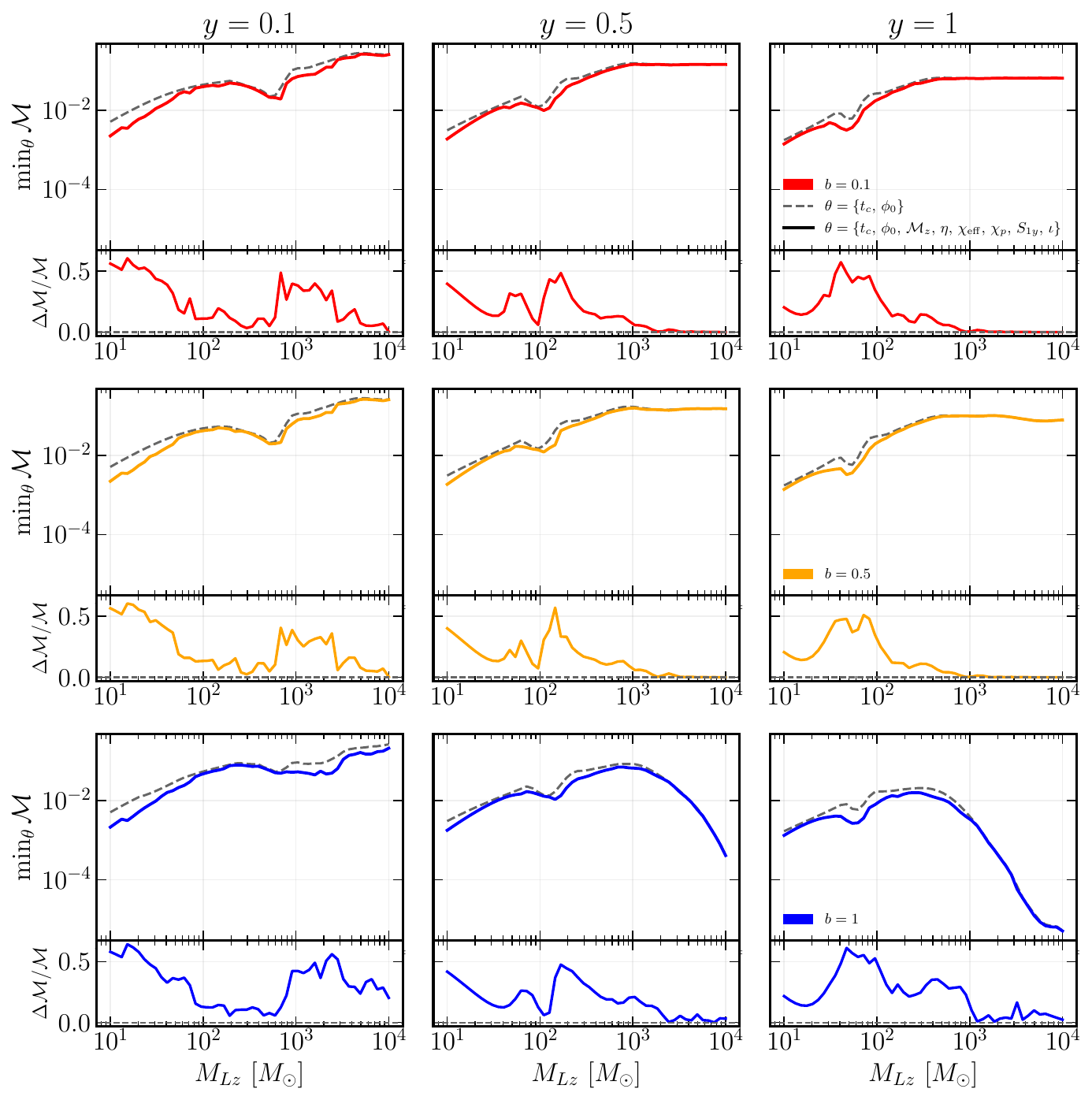}
\caption{\label{fig:mm_scheme_grids_plot_ratio_b} Minimized mismatch assuming $y = 0.1, 0.5,$ and $1$ across the columns, and $b = 0.1, 0.5,$ and $1$ across the rows. In each panel, the dashed curves represent minimization over two parameters $t_c$ and $\phi_0$, whereas the solid curves are minimizations of the mismatch over eight GW waveform parameters, $\theta = \{ t_c, \phi_0, \mathcal{M}_z, \eta, \chi_{\rm eff}, \chi_p, S_{1y}, \iota \}$. Relative residuals between the different minimization schemes are also plotted in each panel.}
\end{figure}

In the main text we significantly simplified the comparison between lensed and unlensed waveforms by carrying out the mismatch minimization in \cref{eq:mismatch} only over the coalescence phase $\phi_0$ and time $t_c$. As a result, our sensitivity estimates are somewhat optimistic, since they effectively rely on the assumption that variations in the remaining parameters (masses of the merging objects, their spins, orientation of the orbit and source distance for fixed source location) \emph{cannot} absorb the diffractive lensing effects. As discussed in the main text, lensing imparts amplitude and phase modulations onto the underlying waveforms. While the amplitude is degenerate with the source distance, the mismatch is not sensitive to this, so we do not consider minimization with respect to $D_S$ below. The masses (expressed as the redshifted chirp mass $\mathcal{M}_z$ and $\eta = m_1m_2/(m_1+m_2)^2$) do affect both the amplitude and the phase as can be seen in the analytic waveform of \cite{Cutler:1994ys}. Misalignment of the black hole spins $\vec{S}_i$ with the orbital angular momentum can lead to precession, which also produces amplitude modulations of the waveform; the size of this effect depends on the orbital inclination angle $\iota$~\cite{Biscoveanu:2026ikx}.

In this appendix, we perform mismatch minimization in the aforementioned parameters over several slices in the lens parameter space. 
The parameter choices below are motivated by Ref.~\cite{Mishra:2023ddt}, which studied the degeneracies between all of the intrinsic and extrinsic parameters and lensing effects. 
We will find that our approximate mismatch values can be optimistic by at most a factor of $\sim 2$, which translates into a similar uncertainty on the derived DM lens abundance. We also consider how our results change for a more restrictive false positive probability, which implies a higher detection threshold and therefore lower efficiency via \cref{eq:pdet_from_pfp}. Both of these effects are similar in magnitude to different prior choices in the full Bayesian analysis (see \cite{LIGOScientific:2021izm} and \cref{fig:ligo_validation} in the next section).

In \cref{fig:mm_scheme_grids_plot_ratio_b} we show $\min_\theta \mathcal{M}$ for the uniform ball lens for two different minimization schemes: the grey dashed lines minimize only over $\theta = \{ t_c, \phi_0 \}$ as in the main text, while the coloured lines represent minimization over eight GW waveform parameters $\theta = \{ t_c, \phi_0, \mathcal{M}_z, \eta, \chi_{\rm eff}, \chi_p, S_{1y}, \iota \}$, which are expected to have a qualitatively similar effect to lensing as described above; here $\chi_{\rm eff}$ is the effective spin parameter (the mass-weighted projection of total spin onto the angular momentum direction -- see, e.g., \cite{Ng:2018neg} and references therein), and $\chi_p$ is the effective precession spin parameter~\cite{Schmidt:2014iyl,Biscoveanu:2026ikx}. Each panel corresponds to a different combination of impact parameter $y$ and lens radius $b$. The relative differences in the mismatch for these minimization schemes are shown in the bottom panel of each plot. We see that the different minimization schemes give rise to similar minimized mismatches, with the minimization over eight parameters always being lower than the minimization over two parameters. The highest relative difference is $\sim 50\%$. Given that the minimization over eight parameters is much more computationally expensive than the minimization over two parameters, in this work we have used the two-parameter minimization scheme. 

\begin{figure}
\includegraphics[width=0.6\columnwidth]{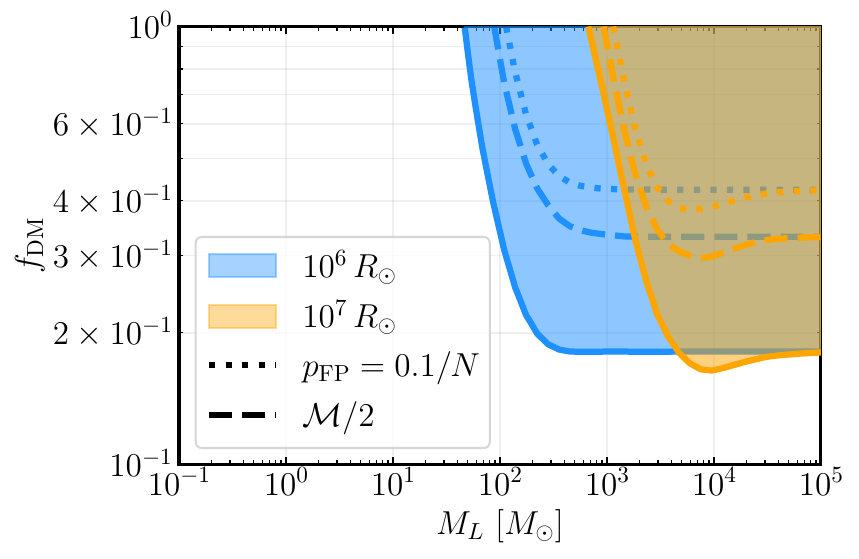}
\caption{\label{fig:f_DM_constraints_LIGO_conservative} Comparison between our primary result presented in the main text (solid) and more conservative estimates for constraints, assuming a lower false positive rate of $p_{\rm FP} = 0.1/N$ (dotted) or a 50\% reduction in the mismatch due to ambiguities in the black hole spin (dashed). }
\end{figure}

We illustrate the impact of reducing the mismatch by a factor of 2 on the LVK sensitivity estimates in \cref{fig:f_DM_constraints_LIGO_conservative} as dashed curves, and compare them to the simplified minimization scheme used in the main text (solid curves). This results in roughly a factor of 2 decrease in the sensitivity to $f_{\rm DM}$. 

In \cref{fig:f_DM_constraints_LIGO_conservative} we also consider the effect of taking a stricter false positive probability $p_{\rm FP}$: the sensitivity with $p_{\rm FP} = 1/N$ (as used in the main text), where $N$ is the number of GW events in the catalogue, is shown as solid curves, while the sensitivity with $p_{\rm FP} = 0.1/N$ is shown as dotted curves. Requiring a lower false positive probability results in a higher threshold for detection, leading to the decrease in sensitivity for $f_{\rm DM}$.

\section{Comparison of point lens constraints with LVK O3}
\label{sec:ligo_validation}
\begin{figure}
\includegraphics[width=0.6\columnwidth]{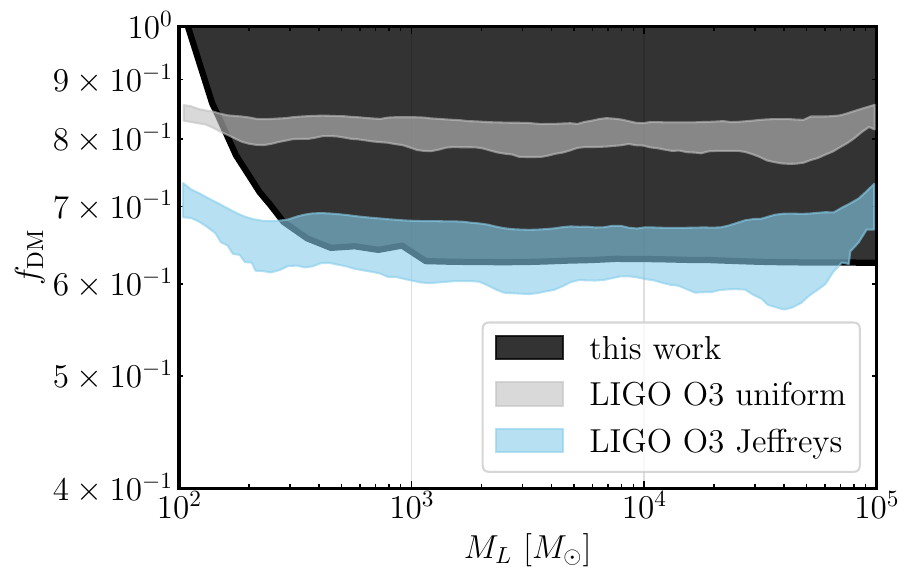}
\caption{\label{fig:ligo_validation} Comparison between the constraints on the point lens masses and fractions obtained through the mismatch calculation in this work and the constraints obtained by the LVK Collaboration from the third observing run O3 \cite{LIGOScientific:2023bwz}. The shaded black region is the constraint obtained using the mismatch in this work. The light blue (grey) regions represent the possible upper bounds from LVK O3 assuming a Jeffreys (uniform) prior in the merger rate modelling. The LVK constraints are shown in a band to indicate different assumptions for the black hole merger rate.}
\end{figure}

As discussed in the main text, the advantage of our semi-analytic statistical treatment is that it provides a quick and simple estimate of the deviation between lensed and unlensed waveforms, which is only expected to be valid for a small amount of deformation. We verify that our constraints for the point lens using O3 data agree with the constraints reported by the LVK Collaboration \cite{LIGOScientific:2023bwz}. We verify against the O3 result because the lensing analysis in GWTC-4.0 did not produce similar plots. Since their analysis did not include any GW events with neutron stars and filtered out source masses smaller than $3 M_\odot$, we perform a similar cut by requiring that $m_1 > 3 M_\odot$ and $m_2 > 3 M_\odot$ in the GWTC-2.1 and GWTC-3 catalogues. In \cref{fig:ligo_validation} we compare our constraints, computed using only the GW events in O3a and O3b, with those of Ref.~\cite{LIGOScientific:2023bwz}. We show their results for both a Jeffreys and a uniform prior on the merger rate. Our results are less sensitive to point lenses with small masses $(\lesssim 400 M_\odot)$ than the result obtained by the LVK Collaboration. Otherwise, the constraints seem to be consistent throughout most of the mass range, except we obtain a drop-off in sensitivity at larger masses than the LVK results. Nevertheless, for the majority of the lens masses our results remain conservative compared to the LVK results from running a full statistical analysis. 

\section{Icarus constraints on uniform density lenses}
\label{sec:icarus_uniform}

In this Appendix, we extend the analysis in Ref.~\cite{Croon:2025yfj} to uniform density lenses. The projected mass for a uniform ball is given by \cite{Croon:2020wpr}
\begin{align}
  m(t) = 
  \begin{cases}
    1 - (1 - t^2/t_m^2)^{3/2}, & |t| < t_m, \\
    1, & |t| \geq t_m ,
  \end{cases}
\end{align}
where $t_m = R/R_E$. The effective Einstein radius in a macrolensing environment is given by $\bar{R}_{E, \rm EDO} \equiv \epsilon_{\rm EDO} \sqrt{\mu_t} \theta_E D_L$, where $\mu_t$ is the tangential macro-magnification of the star and the efficiency factor is given by
\begin{align}
  \epsilon_{\rm EDO}^2 = m(\epsilon_{\rm EDO} \sqrt{\mu_t}). \label{eq:eps_EDO}
\end{align}
The observed Icarus peak magnitude $a_{\rm peak, obs} \sim 26$ is then compared with the minimum peak magnitude, which is given by
\begin{align}
  a_{\rm peak, min} \approx 25.1 - 0.625 \log_{10} \left( \frac{M_{\rm lens}}{M_\odot} \right) - 1.875 \log_{10} \left( \frac{\mu_t}{100} \right),
\end{align}
where $M_{\rm lens}$ is the mass of the lens. Therefore the required $\mu_t$ is given by
\begin{align}
  \mu_{t, \rm req} \approx 33.11 \left( \frac{M_{\rm lens}}{M_\odot} \right)^{-1/3}. \label{eq:mu_req}
\end{align}
\cref{eq:eps_EDO,eq:mu_req} can then be used together to find $\epsilon_{\rm EDO} (M,R)$. The authors of Ref.~\cite{Croon:2025yfj} divide their analysis into two parts: one where $M_{\rm lens} \geq 2 M_\odot$, in which the lensing observable is due to DM substructure, and one where $M_{\rm lens} < 2 M_\odot$, where they assume the lensing is due to an intracluster star of mass $2M_\odot$. In the large-$M$ regime, $M_{\rm lens}$ is given by the DM ball mass $M$, and the effective Einstein radii of the DM lenses start to overlap when the optical depth $\Theta = \Sigma\pi \bar{R}_E^2/M$ is of order unity, which leads to the saturation of the magnification at
\begin{align}
  \mu_{t, \rm sat, EDO} = \frac{M}{\pi \Sigma \bar{R}_E^2},
  \label{eq:mu_sat}
\end{align}
where $\Sigma = f_{\rm DM} \Sigma_{\rm tot}$ is the surface mass density of the lens and $\Sigma_{\rm tot}$ is the total surface mass density. 
Thus, requiring that this saturated magnification exceeds \cref{eq:mu_req} constrains the lens mass and abundance parameter space. 
Concretely, equating \cref{eq:mu_req} and \cref{eq:mu_sat} gives the maximum allowed fraction of DM 
\begin{align}
  f_{\rm DM} = \frac{1}{\kappa \mu_{t, \rm req} (M) \epsilon_{\rm EDO}^2 (M,R)},
\end{align}
where $\kappa = \Sigma_{\rm tot}/\Sigma_{\rm crit} = 0.83 $ and $\Sigma_{\rm crit} = M/(\pi R_E^2) = c^2 D_S/(4\pi G D_L D_{LS})$. On the other hand, when $M_{\rm lens} < 2 M_\odot$, $M_{\rm lens}$ is set to $2 M_\odot$, and the saturation condition is instead given by $\mu_{t, \rm sat} = \kappa^{-1} f_{\rm DM}^{-1}$. This results in a horizontal floor of $f_{\rm DM, max} \approx 0.05$. 

The constraining power falls off at lower and higher masses. At lower lens masses, the constraints fall off because the microcaustic size becomes smaller than the source, and so Ref. \cite{Croon:2025yfj} sets a conservative limit where constraints are no longer possible, given by $\theta_E / \sqrt{\mu_t} < \beta_R$, where $\beta_R = 2.7 \times 10^{-12} (R_{\rm source}/R_\odot)$ with $R_{\rm source} = 260 R_\odot$. Meanwhile, at the high-mass end, the constraints fall off because $\mu_{t, \rm req}$ increases as the lens mass increases. Interestingly, the constraints for the uniform density sphere are similar in behaviour to the constraints for the boson star as studied in Ref.~\cite{Croon:2025yfj}. We direct interested readers to the original paper for more information.

\twocolumngrid
\bibliographystyle{apsrev4-1}
\bibliography{refs}

\end{document}